\documentclass[aps,prl,twocolumn,superscriptaddress,10pt]{revtex4-2}
\usepackage[final]{graphicx}
\usepackage{sidecap}
\usepackage{wrapfig}
\usepackage[usenames,dvipsnames]{xcolor}
\usepackage{multirow}
\usepackage{mathtools}
\usepackage{fancyhdr}
\usepackage{braket}
\usepackage{cprotect}
\usepackage{tabularx}
\usepackage{empheq}
\usepackage{amsmath}
\usepackage{color, colortbl}
\usepackage[para]{footmisc}
\usepackage{footnote}
\usepackage[normalem]{ulem}
\usepackage{natbib}
\usepackage{stackengine}
\newcolumntype{M}[1]{>{\centering\arraybackslash}m{#1}}
\newcommand\xrowht[2][0]{\addstackgap[.5\dimexpr#2\relax]{\vphantom{#1}}}

\usepackage{hyperref}
\hypersetup{colorlinks=true, linkcolor=blue, citecolor=blue, urlcolor=blue}

\begin{document}
\widetext
\title{Magnetic phase diagram of rare-earth orthorhombic perovskite oxides}

\author{Alireza Sasani}
\affiliation{Physique Th\'eorique des Mat\'eriaux, QMAT, CESAM, Universit\'e de Li\`ege, B-4000 Sart-Tilman, Belgium}

\author{Jorge I\~niguez}
\affiliation{MaterialsResearch and Technology Department, Luxembourg Institute of Science and Technology (LIST), 5 avenue des Hauts-Fourneaux, L-4362, Esch/Alzette, Luxemburg}
\affiliation{Department of Physics and Materials Science, University of Luxembourg, Rue du Brill 41, L-4422 Belvaux, Luxembourg}

\author{Eric Bousquet}
\affiliation{Physique Th\'eorique des Mat\'eriaux, QMAT, CESAM, Universit\'e de Li\`ege, B-4000 Sart-Tilman, Belgium}

\begin{abstract}
Spin reorientation and magnetisation reversal are two important features of the rare-earth orthorhombic provskites ($RM$O$_{3}$'s) that have attracted a lot of attention, though their exact microscopic origin has eluded researchers.
Here, using density functional theory and classical atomistic spin dynamics we build a general Heisenberg magnetic model that allows to explore the whole phase diagram of the chromite and ferrite compounds and to scrutinize the microscopic mechanism responsible for spin reorientations and magnetisation reversals.
We show that the occurrence of a magnetization reversal transition depends on the relative strength and sign of two interactions between rare-earth and transition-metal atoms: superexchange and Dzyaloshinsky-Moriya.
We also conclude that the presence of a smooth spin reorientation transition between the so-called $\Gamma_4$ and the $\Gamma_2$ phases through a coexisting region, and the temperature range in which it occurs, depends on subtle balance of metal--metal (superexchange and Dzyaloshinsky-Moriya) and metal--rare-earth (Dzyaloshinsky-Moriya) couplings.
In particular, we show that the intermediate coexistence region occurs because the spin sublattices rotate at different rates.
\end{abstract}

\maketitle

\section{introduction}
\begin{figure}[htb!]
	\centering
	\includegraphics[width=1\linewidth ,keepaspectratio=true]{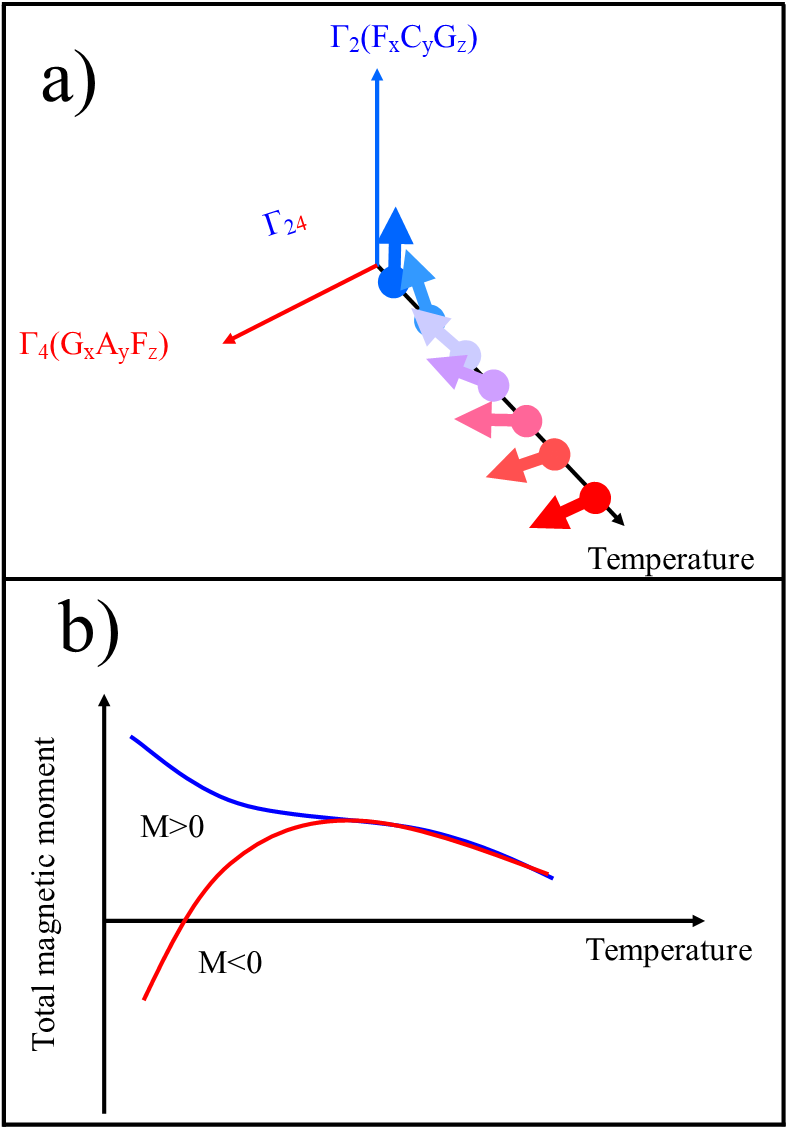} 
	\caption{a) Schematic representation of the SR from $\Gamma_4$ (red color) to $\Gamma_2$ (blue color) as a function of temperature where the transition is smooth by passing through an intermediate mixed phase containing both states ($\Gamma_{24}$). b) Schematic plot of the evolution of the total magnetization of the crystal and showing two possible cases: (i) MR effect (red line) where the magnetization changes sign below a critical temperature due to the fact that the paramagnetic rare earth atom magnetizes in opposite direction to the wFM of the transition metal atom. (ii) Absence of MR (blue line) where the magnetization is amplified when temperature is reduced and corresponding to the case where the rare earth atom magnetizes in the same direction as to the wFM of the transition metal atom. 
	 }
	\label{fig:SR-Mr}
\end{figure}

Rare earth orthorhombic perovskites ($RM$O$_3$, where $R$ is an atom of the rare-earth family and $M$ is a transition metal -- Fe or Cr in this work) have been studied for a long time due to their unique magnetic properties \cite{bousquet2016}, the two important magnetic behaviours being the spin reorientation (SR) and the magnetisation reversal (MR).
The SR involves the change of the spin direction from one crystalline direction to another as a function of temperature (see Fig.~\ref{fig:SR-Mr}(a)) while MR refers to the inversion of the net magnetization of the crystal as a function of temperature (see Fig.~\ref{fig:SR-Mr}(b)).
These materials are also multiferroics (type II, i.e. the magnetic order induces a polarization) \cite{Tokunaga-2008,Tokunaga-2009} with strong magnetoelectric (ME) response \cite{Tokunaga-2008} surpassing most known ME materials.
All of these unique properties rely on the presence of two magnetic sublattices, $R$ and $M$ with very different N\'eel temperatures such that for a wide range of temperatures the $R$ spins are paramagnetic while the $M$ spins are ordered. 
The associated magnetic interactions between these two sublattices have been proved to be the key ingredients for the origin of the SR, MR and multiferroic properties~\cite{Treves-1965,Bazaliy-2004,Yamaguchi1974}, hence for their use in technological applications~\cite{Zhao-2016,Bellaiche-2012,Kimel-2004}. The SR can happen at high temperatures (480~K in the case of SmFeO$_3$) and this temperature can be lowered by doping which makes it possible to have this behaviour at room temperature so that the SR could be used in exchange bias devices~\cite{Kang-2017,GORODETSKY196967, Kimel-2004, Skumryev_2003}.

 Magnetic structures of these materials have been determined from symmetry analysis~\cite{Treves1962}. In this work we are going to use Bertaut notation for symmetry adapted magnetic structures, namely, $\Gamma_1$, $\Gamma_2$, $\Gamma_3$ and $\Gamma_4$ (see Fig.~\ref{fig:Mag-ordr})~\cite{Bertaut}.
Two types of SR are observed, namely $\Gamma_4$ to $\Gamma_2$ (PrFeO$_3$, NdFeO$_3$, SmFeO$_3$, TbFeO$_3$, HoFeO$_3$, ErFeO$_3$, TmFeO$_3$, YbFeO$_3$) and $\Gamma_4$ to $\Gamma_1$ (CeFeO$_3$, DyFeO$_3$)~\cite{bousquet2016}.

During the $\Gamma_4$ to $\Gamma_2$ SR, the spins directions change from the $a$ crystallographic direction to the $c$ direction, slowly rotating as a function of temperature in $ac$ plane.
The $\Gamma_4$ to $\Gamma_2$ SR can develop at different speeds: for some materials it is rather fast (e.g. it spans through a 3~K temperature range \cite{Tsymbal-2007} for ErFeO$_3$) while for others it can occur over a large temperature range (e.g. 70K for NdFeO$_3$ \cite{Constable_2014}). 
Tsymbal {\it et al}. have shown that a mean field model can describe the $\Gamma_4$ to $\Gamma_2$ SR and
they observe a sudden jump at the start of the reorientation and a smooth evolution afterwards \cite{Tsymbal-2007} .
Studies on TbFeO$_3$ show that there are two phase transitions, from  $\Gamma_4$ to $\Gamma_2$ below 8.5~K and, then, at the ordering of Tb the Fe subsystem transforms back to $\Gamma_4$ \cite{Cao-2016} which shows the importance of the $R$ site ordering in this SR. 
It has also been shown that this SR is of second order \cite{Shane-1968,Horner-1968} and it could be associated to the softening of a low frequency magnon mode \cite{White-1982,Scott-1974}.
Bazaliy {\it et al.} have measured the magnetisation of Er in ErFeO$_3$ in the SR region and shown that there is 70\% of change \cite{Bazaliy-2004}.

In the $\Gamma_4$ to $\Gamma_1$ transition, the SR is quite fast and the spins change their directions from $a$ to $b$ crystallographic direction sharply at a well defined transition temperature. It has been shown that the  anisotropic symmetric exchange interactions would be responsible for this SR transition \cite{zvezdin-1979}.

Several authors built models to describe the magnetic interactions in these materials. Moskvin \cite{moskvin1975} has made a model to describe the magnetic interactions in $R$FeO$_{3}$ and $R$CrO$_{3}$ and has also made use of Anderson exchange model \cite{Anderson-1959} to study exchange and DMI in these materials \cite{moskvin1977}.
Yamaguchi~\cite{Yamaguchi1974} used a mean field theory model and was more complete than Moskvin. 
He has shown that some of the SR present in these materials can be explained by antisymmetric and  anisotropic symmetric exchange interactions between $R$ and Fe.
More recently, Bellaiche {\it et al}. have given some simple laws to explain the origin of different cantings present in these structures \cite{Bellaiche-2012}.  In their work they have shown that all the cantings on transition metals site can be described using simple energy terms originating from the DMI between transition metals.
Regarding the interactions between $M$ and $R$ spins, Zhao {\it et al}. have shown that the DMI between $M$ and $R$ can polarize the $R$ ion and hence could explain the origin of the MR in these materials~\cite{Zhao-2016}; according to this work, the MR is linked with a sign in the interaction between the two sublattices that is material dependent and not fully understood.

Hence, although there has been a great effort to explain the magnetic properties of these materials, a solid and complete understanding regarding SR and MR in these materials and their origin is still missing. 

In this article, we shed some light into the magnetic properties of the $RM$O$_3$s. We have used density functional theory (DFT) to fit a microscopic Heisenberg model that includes the superexchange and the DMI interactions between the magnetic cations $M$-$M$ and $M$-$R$ (where $M$ is $Cr$ or $Fe$, and $R$ is Gd).  
This model is then used as starting point, and we tune the different parameters to understand their specific role in magnetic bahaviours of the material using classical spin dynamics.
The spin dynamics results are also compared with analytical solutions to confirm their consistency. 
Our work allows to explain the origin of the SR and the parameters determining the SR temperature interval and how the $R$ magnetism is affected while in its paramagnetic regime.
We find that the occurrence of a slow SR comes from an original evolution of the $\Gamma_4$ and $\Gamma_2$ orders due to the presence of two different interacting magnetic cations; this allows to have two magnetic phases co-existing while no coupling exist between them in the Hamiltonian.

\section{Technical details}
The main goal of this paper is to give a qualitative picture of the magnetic properties of $RM$O$_3$'s. 
To understand these magnetic behaviours we have used DFT calculations on GdFeO$_3$ and GdCrO$_3$, as reference materials, to have an estimation of the order of magnitude of the magnetic interactions in these crystals. We then tuned these parameters to study how they affect the overall magnetic behaviour of the system. 
We build a Heisenberg model containing $M$-$M$ and $M$-$R$ superexchange and DMI interactions.
Because we will focus on the temperature range where the $R$ sublattice is paramagnetic, we will neglect the $R$-$R$ interactions (these interactions are nevertheless small as compared to the $M$-$M$ and $M$-$R$ couplings).
We fit this model against DFT calculations~\cite{Hohenberg1964,Kohn1965} done for the orthorhombic $Pnma$ phase of GdFeO$_3$ and GdCrO$_3$.
We used the VASP package~\cite{KRESSE199615,Kresse-1996} and its projected augmented wave implementation of DFT  \cite{Blochl1994}.
We used the so-called PBEsol-GGA~\cite{Perdew-2008} functional for the exchange correlation part of the density functional; a Hubbard $U$ correction~\cite{Liechtenstein1995} on Fe, Cr and Gd of respectively 4, 2 and 5 eV have been used with $J$ parameter of 1 and 0.5 eV on Fe and Cr. 
All the calculations were done with a 6$\times$6$\times$4 mesh of k-points for sampling the reciprocal space and a cut-off energy on the plane wave expansion of 700 eV to have a good convergence on single ion anisotropic and DMIs (less than 5 $\mu\text{eV}$ convergence).

The calculations of the exchange interactions were done using Green's function method as implemented in the TB2J~\cite{he2020tb2j} code. In this method the maximally localised Wannier function~\cite{Marzari-1997} as implemented in WANNIER90~\cite{MOSTOFI20142309} are calculated using DFT (VASP interface to Maximally localised Wannier functions) and using these Wannier functions and the Green's function method, the exchange parameters are calculated.
Some of these superexchange interactions were compared to the ones calculated using total energy to ensure the consistency of the method. 
To calculate the DMI couplings, we calculated the energy of different spin configuration and used the method given by Xiang {\it et al}.~\cite{xiang_magnetic_2013}.
We have checked that the results are qualitatively the same by using different Hubbard $U$ and $J$ corrections while we have used the ones giving the best N\'eel temperature for both sublattices.
All of the fitted magnetic interaction parameters were used to do spin dynamics with the VAMPIRE code~\cite{evans2014}. In this code the Landau-Lifshitz-Gilbert (LLG) equation for the spin dynamics (Eq.~(\ref{eq:llg})) is solved numerically.

\begin{equation} \label{eq:llg}
\begin{array}{l@{}l}
\frac{\partial S_i}{\partial t}=\frac{\gamma}{1+\lambda^2}\left[S_i\times B_{eff}^i + \lambda S_i \times \left(S_i \times B_{eff}^i \right)  \right]\\
\end{array}
\end{equation}

In the temperature dependent spin dynamics simulations we have used a simulation cell of 20 nanometers in each direction. The thermalisation step was done in 50000 time steps of 1.5~fs and the measurement is done in 90000 time steps of 1.5~fs.

Before analyzing the fitted model, we start with an analytical model of the magnetic interactions present in $RM$O$_3$ systems.

\section{Analytical model}
To understand the mechanism behind SR and MR, we develop in this section the Heisenberg model and solve it analytically to understand the phase diagram of $RM$O$_3$ versus their microscopic magnetic interactions. This will also allow to compare with the spin dynamics calculations to confirm that both give  consistent results.

\subsection{Symmetry adapted spin representation:}
\begin{figure}[htb!]
	\centering
	\includegraphics[width=1\linewidth ,keepaspectratio=true]{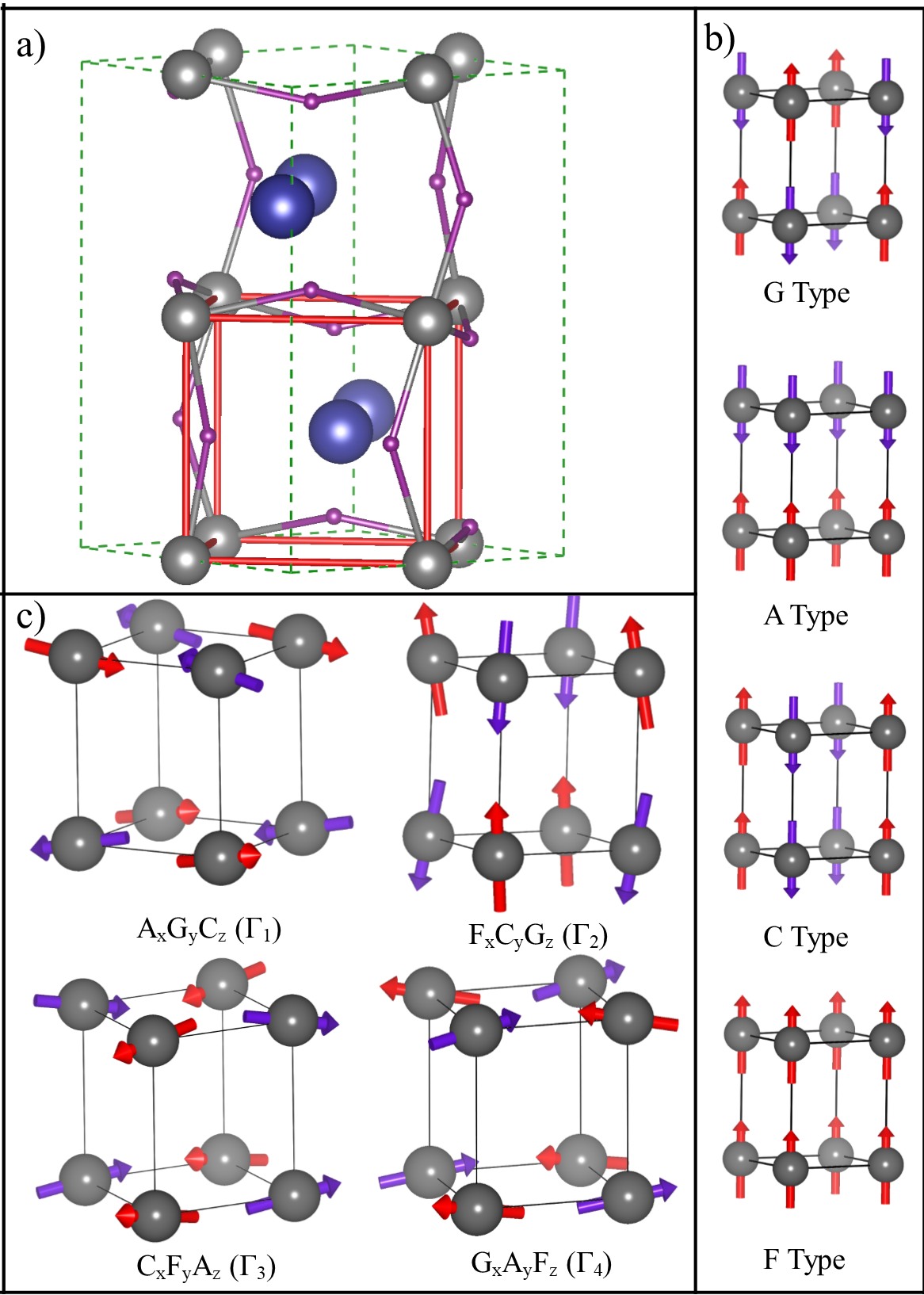} 
	\caption{a) Position of transition metal ions (gray spheres) and rare earth elements (blue spheres) in $Pnma$ structure, purple spheres represent oxygen atoms. b) Schematic representation of the $G$, $A$, $C$ and $F$ magnetic orders for transition metal sites present in $Pnma$ structure as highlighted by the red box in panel a. The arrows represent the positive (red arrows) or negative (purple arrows) value of the magnetic moment. c) Symmetry adapted representations $\Gamma_1$, $\Gamma_2$, $\Gamma_3$ and $\Gamma_4$ present in peroskites.}
	\label{fig:Mag-ordr}
\end{figure}

We develop an analytical model of $RM$O$_3$ using the symmetry adapted spin representation.
For each of the sub-lattices ($M$ or $R$) in the $Pnma$ unit cell, we have four magnetic sites that results in four different magnetic orders: $A$, $C$, $G$ and $F$ type as presented in Fig.~\ref{fig:Mag-ordr}(b). 
Using these four magnetic orderings, we can define four symmetry adapted spin states, namely $\Gamma_1$, $\Gamma_2$, $\Gamma_3$, $\Gamma_4$  that are linear combination of the $A$ ,$G$, $C$ and $F$ orderings in different directions 
(Fig.~\ref{fig:Mag-ordr})\cite{Bertaut}. 
Because the ground state of the $M$ spin sub-lattice is a robust $G$-type antiferromagnetic ordering in the $Pnma$ perovskite phase, the most relevant $\Gamma_j$ states are those with $j$ = 1, 2 and 4, which present a dominant $G$-type in one of the three crystallographic directions with the presence of canted $A$, $C$ and $F$ type magnetic orders in the other directions. 
We summarize in Table \ref{tab:irrep} the different $\Gamma_j$ states where $\bar{G}$ shows the main magnetic order and the components without a bar are small spin cantings.

\begin{table}
 \caption{Irreducible representation of magnetic states present in the $Pnma$ phase of $RM$O$_3$ for both transition metal M-site and $R$ site \cite{Bertaut}}
 \label{tab:irrep}
\centering  
 \begin{tabularx}{\columnwidth}{ >{\setlength\hsize{1\hsize}\centering}X >{\setlength\hsize{1\hsize}\centering}X c}
\hline
Irrep &   M site & $R$ site  \\
\hline
\hline \xrowht{6pt}
 $\Gamma_1$ & (A$_x$,$\bar{G}_y$,C$_z$) & (0,0,C$_z$)  \\
 \xrowht{6pt}
 $\Gamma_2$ & (F$_x$,C$_y$,$\bar{G}_z$) & (F$_x$,C$_y$,0)  \\ 
 \xrowht{6pt}
 $\Gamma_3$ & ($\bar{C}_x$,F$_y$,A$_z$) & (C$_x$,F$_y$,0)  \\ 
 \xrowht{6pt}
 $\Gamma_4$ & ($\bar{G}_x$,A$_y$,F$_z$) & (0,0,F$_z$)  \\ 
\hline
 \end{tabularx}
\end{table}

Using these notations, we can write the symmetry adapted magnetic states in terms of their modulation vectors for magnetic sub-lattice \textit{a} as follows (Eq.~(\ref{eq:1a}) to Eq.~(\ref{eq:1d})):

\begin{widetext}
\begin{equation} \label{eq:1a}
\begin{array}{l@{}l}
S_{i,a}^{\Gamma_1} &{}=A_{a,x}(-1)^{(n_{z}^{i})}+\bar{G}_{a,y}(-1)^{(n_{x}^{i}+n_{y}^{i}+n_{z}^{i})}+C_{a,z}(-1)^{(n_{x}^{i}+n_{y}^{i})}\\
\end{array}
\end{equation}

\begin{equation} \label{eq:1b}
\begin{array}{l@{}l}
S_{i,a}^{\Gamma_2} &{}=F_{a,x}+C_{a,y}(-1)^{(n_{x}^{i}+n_{z}^{i})}+\bar{G}_{a,z}(-1)^{(n_{x}^{i}+n_{y}^{i}+n_{z}^{i})}\\
\end{array}
\end{equation}

\begin{equation} \label{eq:1c}
\begin{array}{l@{}l}
S_{i,a}^{\Gamma_3}=C_{a,x}(-1)^{(n_{y}^{i}+n_{z}^{i})}+F_{a,y}+A_{a,z}(-1)^{(n_{z}^{i})}\\
\end{array}
\end{equation}

\begin{equation} \label{eq:1d}
\begin{split}
S_{i,a}^{\Gamma_4}=\bar{G}_{a,x}(-1)^{(n_{x}^{i}+n_{y}^{i}+n_{z}^{i})}+A_{a,y}(-1)^{(n_{z}^{i})}+F_{a,z}\\
\end{split}
\end{equation}

\end{widetext}

Here $S_{i,a}^{\Gamma_j}$ is the spin of lattice site $i$ for magnetic sub-lattice $a$ ($M$ or $R$) in irreducible representation $\Gamma_j$ and the lattice site vector for lattice site $i$ can be written as $n_{x}^i \hat{u}_1 + n_{y}^i \hat{u}_2 + n_{z}^i \hat{u}_3$ where $\hat{u}_1$,$\hat{u}_2 $ and $\hat{u}_3 $  are unit cell vectors
while the coefficients $G$, $A$, $F$ and $C$ represent the magnitude of spin canting in each direction, the $G$-type order being the main one.
From now on, we will use these spin representations in our Heisenberg model.

\subsection{Heisenberg model}
In this section, we develop the Heisenberg Hamiltonian for $RM$O$_3$ in which we include the magnetic interactions between all the magnetic species: transition metal atoms ($M$) and rare earth atoms ($R$), which can be summarized as follows if one stays at the second order  of  interactions ( higher order spin interactions like biquadratic or four-spin couplings are neglected):
\begin{equation} \label{eq:2}
H=H^{MM}+H^{RM}+H^{RR},
\end{equation}
where $H^{MM}$ is the Hamiltonian of M-M interactions, $H^{RR}$ the Hamiltonian of R-R interactions and $H^{RM}$ the Hamiltonian of R-M interactions.
$H^{MM}$ can be written as follows:
\begin{equation} \label{eq:3}
H^{MM}=H_{ex}^{MM}+H_{DMI}^{MM}+H_{SIA}^{MM},
\end{equation}
where $H_{ex}^{MM}$, $H_{DMI}^{MM}$ and $H_{SIA}^{MM}$ represent the superexchange, DMI and single ion anisotropy (SIA) interactions of the $M$ cations. 
In our simulations we have neglected anisotropic symmetric exchange interactions since our DFT calculations show that they are two orders of magnitude smaller than DMIs (results not shown here).

For $H^{RR}$ we have neglected the $H_{ex}^{RR}$ and $H_{DMI}^{RR}$ since we are interested in behaviours that take place at temperatures higher than the N\'eel temperature of the $R$ spin sub-lattice. 
We will only keep the SIA interactions for this site. 
\begin{equation} \label{eq:5}
H^{RR}=H_{SIA}^{RR}
\end{equation}
The Hamiltonian taking care of the $R$-$M$ interactions can be written as follows:
\begin{equation} \label{eq:4}
H^{RM}=H_{ex}^{RM}+H_{DMI}^{RM}
\end{equation}

The exchange, DMI and SIA terms can be developed as follows:
\begin{equation} \label{eq:6}
H_{Ex}^{ab}=\frac{1}{2}\sum_{ij}^N\left( J_{ab,ij}S_{i,a}.S_{j,b}  \right) \\
\end{equation}
\begin{equation} \label{eq:7}
H^{ab}_{dmi}=\frac{1}{2}\sum_{i,j}\left(D_{ab,ij}\times S_{j,a}\right).S_{i,b}\\
\end{equation}
\begin{equation} \label{eq:8}
H^{aa}_{SIA}=\sum_{i}K_{a}\left(S_{i,a}.\hat{e}_{i}\right)^2,\\
\end{equation}
where $ab$ could be $a=b=M$, $a=b=R$ or $a=M$ and $b=R$ and $\hat{e}_i$ is a unit vector pointing to the direction of the SIA  axis, which, according to our DFT calculation for GdFeO$_3$ and GdCrO$_3$, is the easy axis.


We can show that there is no interaction between the different $\Gamma_j$ magnetic orderings (see supplementary information (SI)) such that we can write the total energy of the system as the sum of the energy of each state. 
In this case we can write the total energy as:

\begin{equation} \label{eq:9}
\begin{split}
H=H^{\Gamma_1}+H^{\Gamma_2}+H^{\Gamma_3}+H^{\Gamma_4}\\
\end{split}
\end{equation}

In our analytical derivations we have neglected the $\Gamma_3$ state since this state is much higher in energy than $\Gamma_1$, $\Gamma_2$ and $\Gamma_4$.
This is related to the fact that the $\Gamma_3$ state does not contain $G$-type order, which is the order driving the lowest energy in the crystal through the strongest superexchange interactions between transition metals.

By putting the spin states in the Hamiltonian we can derive the following expressions for each of the states (see SI):
\begin{widetext}
\begin{equation} \label{eq:10a}
\begin{split}
H^{\Gamma_1} =  &{}H_{ex}^M+H_{DMI}^M+H_{ex}^{RM}+H_{DMI}^{RM}\\
=&{} NJ^M(A_{M,x})^2-3NJ^M(\bar{G}_{M,y})^2-NJ^M(C_{M,z})^2\\
&{} -6N d_x^{M}\bar{G}_{M,y}C_{M,z}-6N d_y^{M}C_{M,z}A_{M,x}-6N d_z^{M}A_{M,x}\bar{G}_{M,y}\\
&{} -8Nd_{x}^{RM}C_{R,z}\bar{G}_{M,y}-8Nd_{y}^{RM}C_{R,z}A_{M,x}
\end{split}
\end{equation}

\begin{equation} \label{eq:10b}
\begin{split}
H^{\Gamma_2}=   &{}H_{ex}^M+H_{DMI}^M+H_{ex}^{RM}+H_{DMI}^{RM}\\
=&{}3NJ^M(F_{M,x})^2-NJ^M(C_{M,y})^2-3NJ^M(\bar{G}_{M,z})^2\\
&{} -6N d_x^{M}C_{M,y}\bar{G}_{M,z}-6N d_y^{M}\bar{G}_{M,z}F_{M,x}-6N d_z^{M}F_{M,x}C_{M,y}\\
&{} -8NJ^{RM}F_{M,x}F_{R,x}-8Nd_{x}^{RM}\bar{G}_{M,z}C_{R,y}-8Nd_{y}^{RM}F_{R,x}\bar{G}_{M,z}\\
&{} -8N d_{z}^{RM}F_{R,x}C_{M,y}-8N d_{z}^{RM}C_{R,y}F_{M,x}
\end{split}
\end{equation}

\begin{equation} \label{eq:10c}
\begin{split}
H^{\Gamma_4}= &{}H_{ex}^M+H_{DMI}^M+H_{ex}^{RM}+H_{DMI}^{RM}+H_{SIA}^R+H_{SIA}^M\\
=&{} -3NJ^M(\bar{G}_{M,x})^2+NJ^M(A_{M,y})^2+3NJ^M(F_{M,z})^2\\
&{} -6Nd_x^{M}A_{M,y}F_{M,z}-6Nd_y^{M}\bar{G}_{M,x}F_{M,z}-6Nd_z^{M}\bar{G}_{M,x}A_{M,y}\\
&{} -8NJ^{RM}F_{M,z}F_{R,z}-8Nd_{x}^{RM}F_{z,R}A_{M,y}-8Nd_{y}^{RM}F_{R,z}\bar{G}_{M,x}\\
&{} -N K^M(\bar{G}_{M,x})^2-N K^R(\bar{G}_{R,x})^2 
\end{split}
\end{equation}
\end{widetext}

Where $J^M$ and $J^{RM}$ are, respectively, the exchange interaction magnitude for $M$ sublattice and between $R$ and $M$ spins(J$^a$ is considered as J$_{ii}$ for isotropic exchange interaction); $d_{i}^{a}$ is the magnitude of $i$th component of DMI vector for magnetic sub-lattice $a$; and $N$ is the number of magnetic atoms while $K^a$ represent the SIA magnitude of magnetic sublattice $a$.

With Eqs.~(\ref{eq:10a}--\ref{eq:10c}) 
we have decomposed the Hamiltonian in terms of three independent representations $\Gamma_1$, $\Gamma_2$ and $\Gamma_4$, themselves decomposed into the superexchange, DMI and SIA of their constituent $A$, $C$, $G$ and $F$ magnetic orderings.
This form allows us to decompose the different microscopic contributions of the magnetic energy of the $RM$O$_3$ systems.

\section{DFT calculation of the magnetic interaction parameters:}
In this section we will present the parameters that we have calculated using DFT for GdFeO$_3$ and GdCrO$_3$, which will serve as a reference starting point in our spin dynamics simulations. 
These values will guide us to scan the magnetic phase diagram in regions that are relevant for these materials.

\subsection{Superexchange and DMI parameters}

\begin{table}
\caption{Calculated magnetic interactions from DFT of GdFeO$_3$ and GdCrO$_3$ (units are meV). $J$ (average values between nearest neighbours) values are for the first nearest neighbours and $d_i$ are the DMI vector components along $i=$ $x$, $y$ and $z$.} 
\label{tab:DMI}
\centering
 \begin{tabularx}{\columnwidth}{ >{\setlength\hsize{1\hsize}\centering}X >{\setlength\hsize{1\hsize}\centering}X >{\setlength\hsize{1\hsize}\centering}X >{\setlength\hsize{1\hsize}\centering}X  c}
\hline
\hline
     & d$_x$ & d$_y$ & d$_z$ & J\\
\hline \xrowht{6pt}
Fe-Fe & 0.000  & -1.805 & -1.104  &  38   \\
 \xrowht{6pt}
Cr-Cr & -0.001 & -0.810 & -0.600  &  7.2  \\
 \xrowht{6pt}
Gd-Fe & 0.008  & -0.064 & 0.031   &  1.85 \\
 \xrowht{6pt}
Gd-Cr & -0.010 &  0.042 & -0.019  &  2.15 \\
 \xrowht{6pt}
Gd-Gd & --     & --     & --      &  0.19 \\
\hline
\end{tabularx}
\end{table}

The dominant interactions are the exchange interaction between transitions metals. The DFT results for GdFeO$_3$ and GdCrO$_3$ show that the strongest exchange interactions are between the nearest neighbours transition metals; going further in distance gives very small values with respect to the nearest neighbours such that they can be neglected. 
These interactions are 38 meV and 7.2 meV for nearest neighbours (see Table \ref{tab:DMI}), in GdFeO$_3$ and GdCrO$_3$ respectively and 1 meV or below for the next nearest neighbours.
The $R$-$M$ superexchange interactions are one order of magnitude smaller (approximately 2 meV) than the ones between transition metals.
The $R$-$R$ superexchange interactions are two orders of magnitude smaller than the transition metals ones (around 0.2 meV) such that we have neglected the $R$-$R$ interactions in our spin dynamics simulations.
Calculated SIA for both sublattices in GdFeO$_3$ shows that these parameters are small around 72 $\mu \text{eV}$ for Gd with easy axis along $c$ direction and 75 $\mu \text{eV}$ for Fe with easy axis along the $b$ in the $Pnma$ structure; in GdCrO$_3$ the SIA constant for Cr in along $c$ direction and it is 25 $\mu \text{eV}$.

\begin{figure}[htb!]
	\centering
	\includegraphics[width=1\linewidth ,keepaspectratio=true]{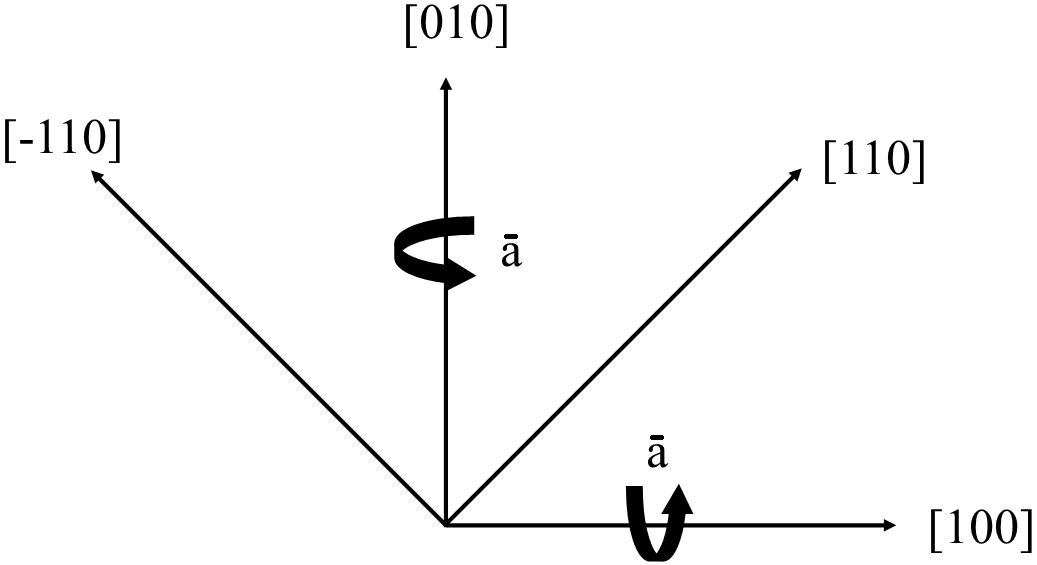} 
	\caption{Schematic presentation of the cubic ([100] and [010]) and $Pnma$ ([110] and [$\bar{1}$10] ) crystallographic directions with respect to each other. The curved arrows represent the oxygen octahedra rotations that are in the same direction when projected in the [110] direction while they are in opposite direction when projected in the [$\bar{1}$10] direction.}
	\label{fig:rot-dire}
\end{figure}

The most relevant parameters for our behaviours of interest are the DMIs.
Table \ref{tab:DMI} shows the obtained results, where the relation $d_y > d_z \gg d_x$  always hold. 
It is the $\overset{-}{a}\overset{-}{a}\overset{+}{c}$ octahedral rotation pattern that breaks the bond inversion center of symmetry and creates the DMI \cite{weingart2012}. 
Hence, we will have the biggest distortion in [110] cubic direction (amplitude of the rotations in $y$ direction of the $Pnma$ structure) and the smallest one will be in the [$\overset{-}{1}$10] cubic direction ($x$ direction of the $Pnma$ structure), (as shown in Fig.~\ref{fig:rot-dire} the oxygen octahedral rotation have the same sign in [110] direction and add up while in [$\overset{-}{1}$10] direction they have opposite sign and subtract from each other), while the distortion in the [001] cubic direction ($c$ direction of $Pnma$) will be almost half of the one in the [110] cubic direction.  
The ratio between these distortions is close to be the same for any $Pnma$ crystal and this structural ratio also drives the key magnetic interactions as we will show below.

At high temperatures there is no magnetic ordering on $R$ sites (paramagnetic phase) and the interactions between the $R$ spins are negligible. 
Hence, at high temperatures the SIA and DMI interactions of the $M$ sites determine the magnetic equilibrium state.
From the formulas \ref{eq:10a},\ref{eq:10b} and \ref{eq:10c} we can notice  that both $d_{y}^{M}$ and $d_{z}^{M}$, which have the biggest components compared to $d_{x}^{M}$ (see table \ref{tab:DMI}) in the $\Gamma_4$ state, are coupled with the main magnetic order and spin direction (i.e $G_{M,x}$), making the energy of this state lower compared to $\Gamma_1$ and $\Gamma_2$. 
When comparing $\Gamma_1$ and $\Gamma_2$, we can observe that for the $\Gamma_2$ state we have the $d_{y}^{M}$ terms that couple with the main spin direction, hence stronger than the $d_{z}^{M}$ component present in $\Gamma_1$. 
This implies that the $\Gamma_2$ state is lower in energy than the $\Gamma_1$ sate. Hence we can have:
\begin{center}
    $E_{DMI}^{MM,\Gamma_4} <$ $E_{DMI}^{MM,\Gamma_2} < $ $E_{DMI}^{MM,\Gamma_1}$
\end{center}
where $E_{DMI}^{MM,\Gamma_j}$ is the energy from DMI between $M$ atoms in the $\Gamma_j$ state, i.e. the DMI between $M$ cations favour the $\Gamma_4$ state \cite{moskvin1975}.

We can also see the effect of these interactions in $R$CrO$_3$ structures. 
According to table \ref{tab:DMI} the DMIs and exchanges in these structures are smaller than for $R$FeO$_3$, which makes the energy difference between different spin orders ($\Gamma_1$, $\Gamma_2$, $\Gamma_4$) smaller. 
This is consistent with the fact that the $\Gamma_2$ and $\Gamma_4$ states are both present at high temperature for the $R$CrO$_3$ crystal series \cite{Shamir-1981}.

Considering the DMIs between $R$ and $M$, we notice that in the $\Gamma_2$ state we have $H^{RM,\Gamma_2}_{DMI}=-8Nd_{x}^{RM}\bar{G}_{M,z}C_{R,y}-8Nd_{y}^{RM}F_{R,x}\bar{G}_{M,z}-8N d_{z}^{RM}F_{R,x}C_{M,y}-8N d_{z}^{RM}C_{R,y}F_{M,x}$ terms in which $d_{y}^{RM}$ couples with the main spin directions and also this state has more degrees of freedom compared to the other states making energy of this state lower. As for $\Gamma_4$  we have $H^{RM,\Gamma_4}_{DMI}=-8Nd_{x}^{RM}F_{z,R}A_{M,y}-8Nd_{y}^{RM}F_{R,z}\bar{G}_{M,x}$ and for $\Gamma_1$ we have $H^{RM,\Gamma_1}_{DMI} =-8Nd_{x}^{RM}C_{R,z}\bar{G}_{M,y}-8Nd_{y}^{RM}C_{R,z}A_{M,x}$ which again due to having the coupling $8Nd_{y}^{RM}F_{R,z}\bar{G}_{M,x}$ compared to $8Nd_{x}^{RM}C_{R,z}\bar{G}_{M,y}$ terms the $\Gamma_4$ state is lower than that of $\Gamma_{1}$ ($d_{y}^{RM} \gg d_{x}^{RM}$).
Hence we can write the order of the different energies due to DMI of $R$ and $M$ as:\\
\begin{center}
 $E_{DMI}^{RM,\Gamma_2} <$ $E_{DMI}^{RM,\Gamma_4} < $ $E_{DMI}^{RM,\Gamma_1} $.
\end{center}
 From this analysis, we can see why a SR transition is possible when lowering the temperature.

Indeed, as the temperature is lowered the interactions between $R$ and $M$ cations become stronger due to the magnetization of the $R$ site in the field created by the $M$ spins, and the $\Gamma_2$ is more and more favoured through the DMIs between $R$ and $M$ sites.
Hence, we can explain the $\Gamma_4$ to $\Gamma_2$ SR due to the DMIs between $R$ and $M$ sites as discussed previously by Yamaguchi \cite{Yamaguchi1974}.

\subsection{Origin of ordering on $R$ site}

The MR observed in $RM$O$_3$ is the change of sign in the net magnetization of the material.
This property has been related to the polarisation of the $R$ site atoms as a result of interaction with transition metal atoms. In this interaction $R$ site atoms could polarise in the direction of the weak magnetic moment of the transition metal or in opposite direction, which would result in the presence or the absence of the magnetisation reversal respectively (see Fig.~\ref{fig:SR-Mr}(b)). 
The remaining question is why the $R$ paramagnetic atoms magnetize in opposite direction for some $R$ elements (e.g. NdFeO$_3$, SmFeO$_3$, DyFeO$_3$, ErFeO$_3$, TmFeO$_3$, YbFeO$_3$) and why they magnetize in the same direction for others (e.g. PrFeO$_3$, EuFeO$_3$, GdFeO$_3$, TbFeO$_3$, HoFeO$_3$) \cite{ZHOU201421} and what is the microscopic origin of this effect.

With our model we can have access to the detailed microscopic interaction between $R$ and $M$ cations. 
Equations \ref{eq:10a},\ref{eq:10b} and \ref{eq:10c} show that there are two types of interactions acting on $R$ sites: (i) the exchange interaction between the weak ferromagnetic (wFM) order of the $M$ and $R$ sublattices ($J_{RM}F_{M}F_{R}$) and (ii) the DMI between the $G$-type orders on the $M$ and $R$ sublattices ($d_{y}^{RM}F_{R,(z,x)}\bar{G}_{M,(x,z)}$ and $d_{x}^{RM}C_{R,z}\bar{G}_{M,y}$).
These interactions can induce either $F$ or $C$ type ordering on $R$ site.
To check the validity of these possibilities, we have used DFT calculations as computer experiments where we have replaced the $R$ site by Cr in GdFeO$_3$'s $Pnma$ structure to allow the study of full non-collinear calculations and to have stronger $RM$ site interactions compared to the Gd case. 
Our model is valid for two magnetic sublattices in perovskites whatever the magnetic cations, such that replacing $R$ by Cr will show the same qualitative trend.

\begin{figure}[htb!]
	\centering
	\includegraphics[width=1\linewidth ,keepaspectratio=true]{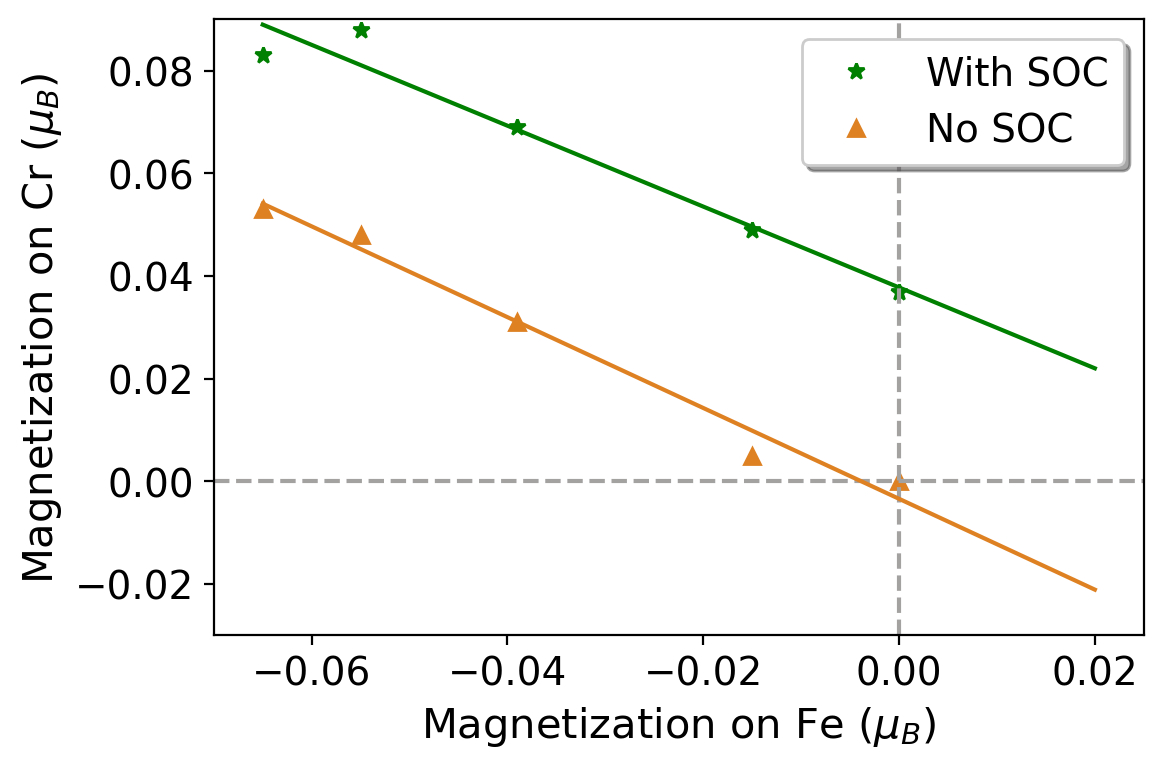} 
	\caption{Calculated magnetization of Cr as a function of magnetization of Fe in CrFeO$_3$ simulated at fixed atomic positions of relaxed $Pnma$ GdFeO$_3$ (Gd is replaced by Cr). 
	Orange points are without spin orbit coupling (with a linear fit orange line) and green points are with spin orbit coupling (with linear fit green line).}
	\label{fig:Fe-Cr}
\end{figure}

We have done different calculations in which we constrained the magnetic moments on the Fe site and relaxed the magnetic order of the $R$ site within two different settings. 
In the first setting, we set the spin orbit coupling (SOC) to zero to suppress the DMI such that the resulting magnetic order on $R$ site would be due to superexchange interactions only. 
In the second setting, we considered SOC, hence activating the second term ($d_{y}^{RM}F_{R,(z,x)}\bar{G}_{M,(x,z)}$) that couples the $G$ type order of the $M$ sublattice to $F$ order on $R$ site. 
In Fig.~\ref{fig:Fe-Cr} we show the result of these two types of calculations where we can see that the magnetization line jumps to higher values when the SOC is present. 
This shows that the DMI can polarise the $R$ site as the superexchange, which is in agreement with the results obtained by Zhao {\it et al}.~\cite{Zhao-2016} using DMI energetic expressions between $R$ and $M$ sites. 

We should also mention that in our simulations for $R$ = Cr, since the superexchange interaction is AFM, it polarises the $R$ site in opposite direction to the wFM direction of the $M$, while the DMI  polarises the $R$ site in the same direction as the wFM of the $M$ site (see Fig.~\ref{fig:Fe-Cr}). 
Our calculations for GdFeO$_3$ also show that these interactions are in competition with each other such that the final magnetization direction of the $R$ site will be determined by the balance between them.

Considering GdCrO$_3$ we can see in Table~\ref{tab:DMI} that the DMI interactions between the Gd and Cr spins has opposite sign compared to the DMI interaction between Gd and Fe spins. 
This shows that we can also have the sign change of the DMI depending on the electronic structure of the atoms in the structure. 
In this case both the DMI and superexchange induce a polarisation in the same direction and, indeed, the calculated magnetic ground state of the GdCrO$_3$ shows that the wFM of Cr and polarisation of the Gd are in opposite directions, while for GdFeO$_3$ the wFM of the iron atoms and the polarisation direction of the Gd atoms are in the same direction.
Hence, depending on the electronic structure of the atoms we can have the superexchange and DMI that compete or cooperate that will result in the presence or the absence of the MR.

\section{Spin dynamics}
In this section we present the spin dynamics results obtained with the VAMPIRE code through the Heisenberg model presented above (with $R-R$ superexchange and DMI set to zero).
 First, we worked with the magnetic interactions parameters obtained for GdFeO$_3$. Then, we made additional spin dynamics calculations by varying the values of these parameters to understand how the phase diagram and associated SR transitions are affected by the change of the magnetic interactions.

To verify that our model qualitatively respects the symmetry of the $Pnma$ phase of $RM$O$_3$ compounds, we first simulated the ground state (0~K) of these structures by tuning the SIA to obtain the magnetic moment direction along the different $x$, $y$ and $z$ crystallographic directions. By doing so, we verified that the obtained cantings actually correspond to the ones of the $\Gamma_4$, $\Gamma_1$ and $\Gamma_2$ orders when magnetic moments lie along $x$, $y$ and $z$ directions, respectively (see table \ref{tab:irrep}).

In the following we will analyze both $\Gamma_4$ to $\Gamma_2$ and $\Gamma_4$ to $\Gamma_1$ SR transitions.

\subsection{ $\Gamma_4$ to $\Gamma_2$ reorientation}
\begin{figure}[htb!]
	\centering
	\includegraphics[width=1\linewidth ,keepaspectratio=true]{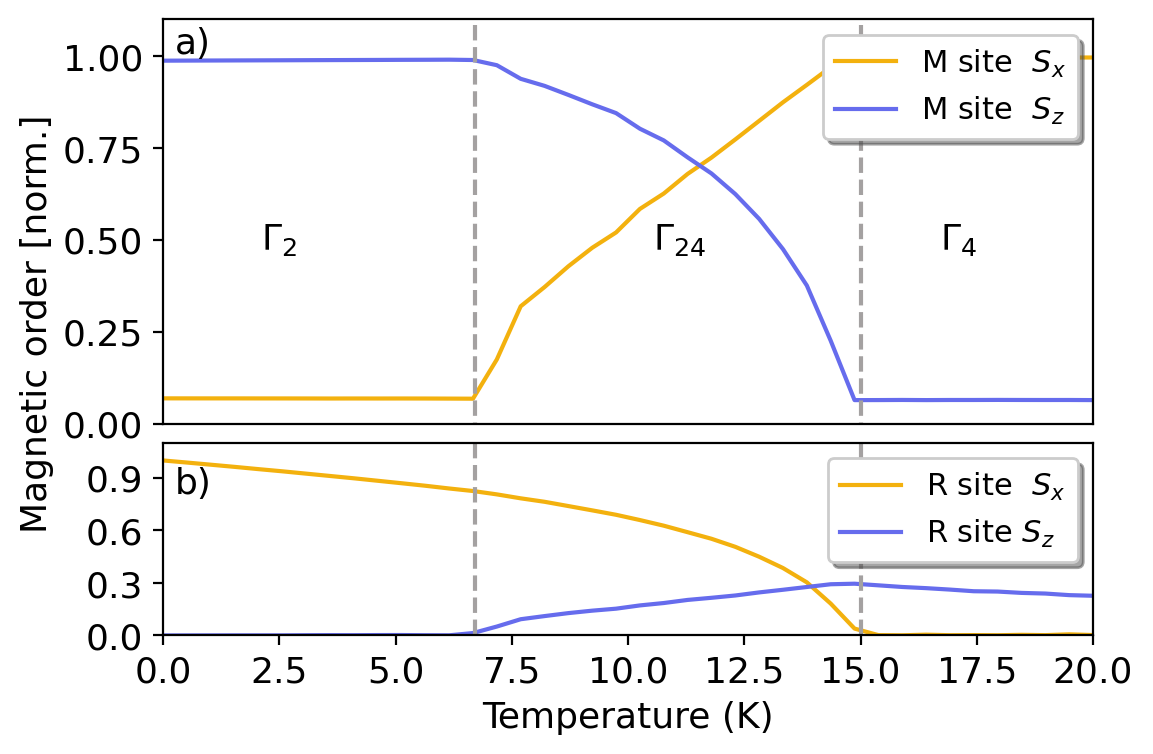} 
	\caption{Temperature dependent SR as obtained from our spin dynamics calculations. Panel (a) shows the transition-metal $M$ site spin projections along the $x$ and $z$ directions. Panel (b) shows the evolution of the $x$ and $z$ magnetic moment projection of the rare earth $R$ site in the same temperature range as panel (a). The unis on y axis are spins normalised with their moments($\frac{5}{2}$ and $\frac{7}{2}$ for Fe and Gd respectively.) }
	\label{fig:T_SR}
\end{figure}

As first step we have done temperature dependent SR. 
To have temperature dependent SR, we have tuned the parameters obtained for GdFeO$_3$ so as to induce such behavior, since this effect is not present in GdFeO$_3$. More precisely, we increase the DMI interaction between $R$ and $M$ by 1 order of magnitude to have the SR. Fig.~\ref{fig:T_SR}(a) shows the evolution of the magnetic moment directions with respect to the temperature when there is a SR, as obtained from our spin dynamics simulations.

We can see a slow rotation of the spins from $x$ to $z$ direction as the temperature decreases and that this reorientation is continuous in a range of temperature where the two orders (associated to the $\Gamma_4$ and $\Gamma_2$ states) are present together.
Fig.~\ref{fig:T_SR}(b) shows the temperature evolution of the magnetic ordering of the $R$ spins due to its interaction with the $M$ spins.
Here, we can see that the SR happens around 15~K when the normalized magnetic moment of the rare earth element is about 0.3. Below this critical temperature, the $R$ site magnetic moment increases in a more pronounced way.
The magnetization of the $R$ sites creates a torque that induces the rotation of the $M$ magnetic direction. 
Such an increase of the ferromagnetic moment of the $R$ site has been observed experimentally for ErFeO$_3$ in the SR region \cite{Bazaliy-2004}. 
This result shows that we need the ferromagnetic ordering on the $R$ site to have this $\Gamma_4$ to $\Gamma_2$ SR transition.
This is also observed experimentally in, e.g., TbFeO$_3$ where the crystal goes from the $\Gamma_2$ state to the $\Gamma_4$ state when the Tb atom orders into the A$_x$G$_y$ magnetic phase (no ferromagnetic order) at very low temperatures \cite{Cao-2016}.

\begin{figure}[htb!]
	\centering
	\includegraphics[width=1\linewidth ,keepaspectratio=true]{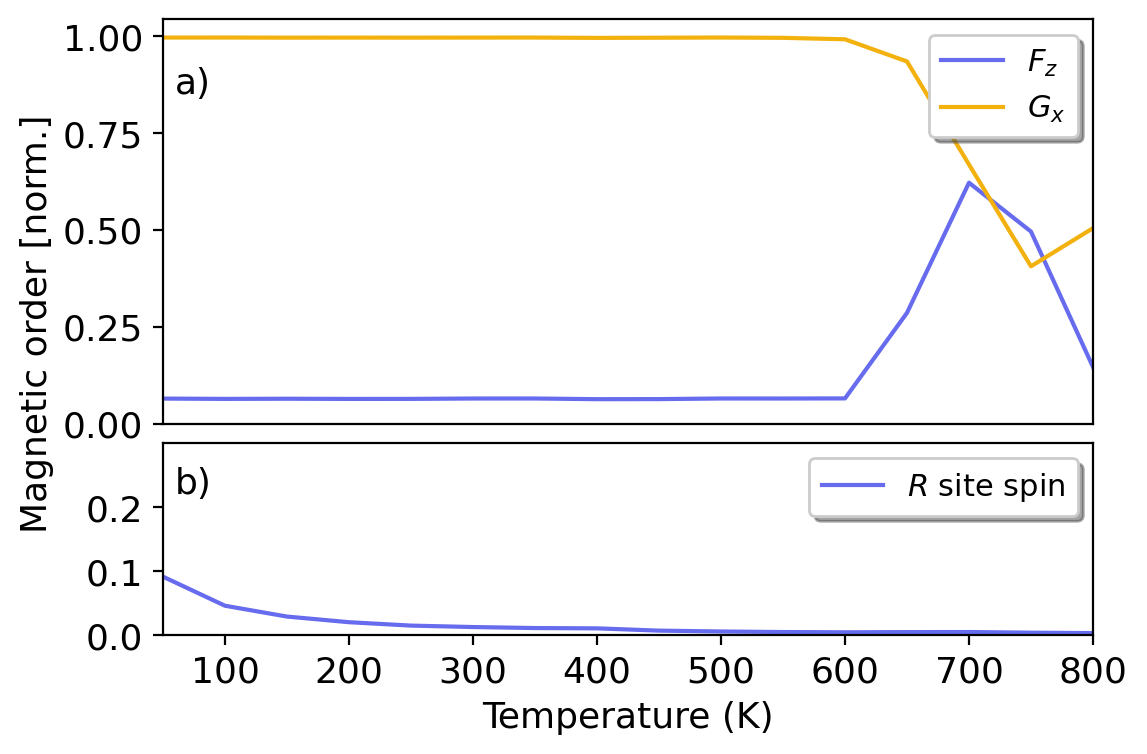} 
	\caption{Temperature dependent spin dynamics results for GdFeO$_3$: a) normalized magnetic moment of the Fe site as projected along z and x direction(F$_z$ and G$_x$) and b) normalized magnetic moment of the Gd site (ferromagnetic order along the $z$ direction). 
	}
	\label{fig:T_SR2}
\end{figure}

Fig.~\ref{fig:T_SR2}(a) shows the evolution of the magnetic orders at higher temperature. 
This figure shows a N\'eel temperature of 680~K (using the $M-M$ superexchange parameters as obtained for GdFeO$_3$) and that the wFM (F$_z$) appears at temperatures below the N\'eel temperature and, after a jump at the phase transition, stays constant (0.10 $\times \frac{5}{2} \mu_{B}$).
In Fig.~\ref{fig:T_SR2}(b) we show the evolution of the $R$ site magnetic moments where we can see that the induced magnetization of the rare-earth spins is visible at temperatures as high as 400~K.

\begin{figure*}[htb!]
	\centering
	\includegraphics[width=\textwidth ,keepaspectratio=true]{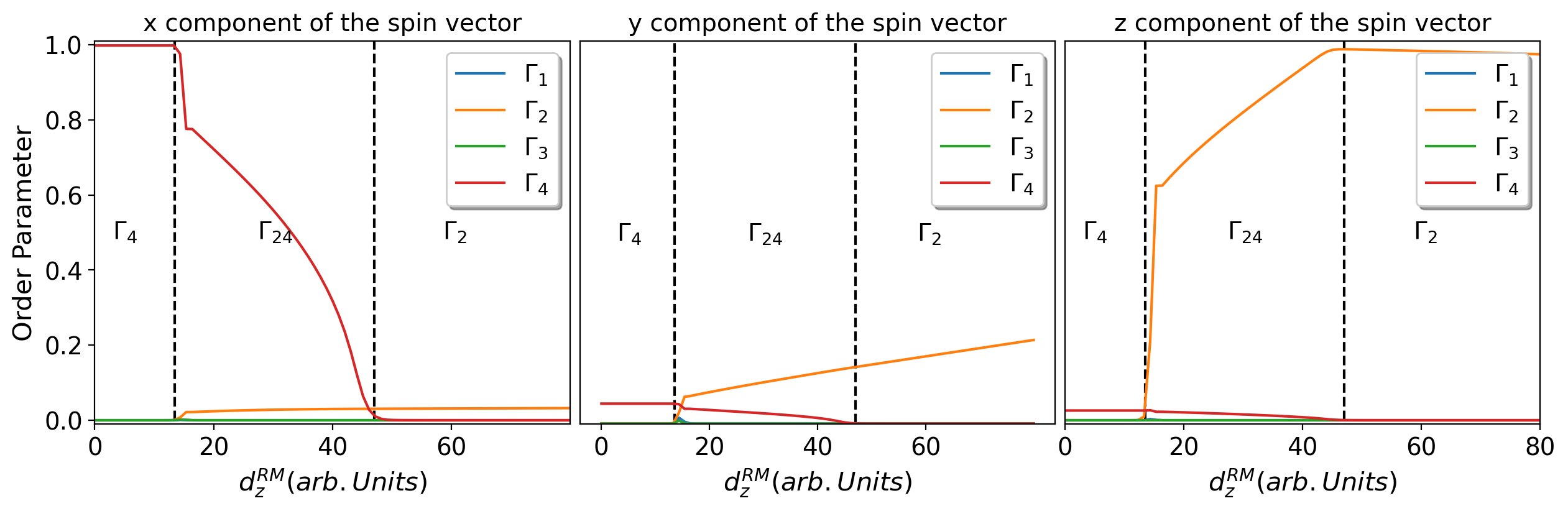} 
	\caption{Decomposition of magnetic ordering on different irreducible representations ($\Gamma_j$) for $x$, $y$ and $z$ directions of the spins in the $\Gamma_4$ to $\Gamma_2$ SR phase transition case. The calculations are done at 0K (ground state). The horizontal axis shows the magnitude of the $z$ component DMI between $R$ and $M$ sites ( we have multiplied the calculated DMI component with these values to increase it)
	and the vertical axis shows the normalised order parameter magnitude.
	}
	\label{fig:Fe_SR}
\end{figure*}

\begin{figure}[htb!]
	\centering
	\includegraphics[width=1\linewidth ,keepaspectratio=true]{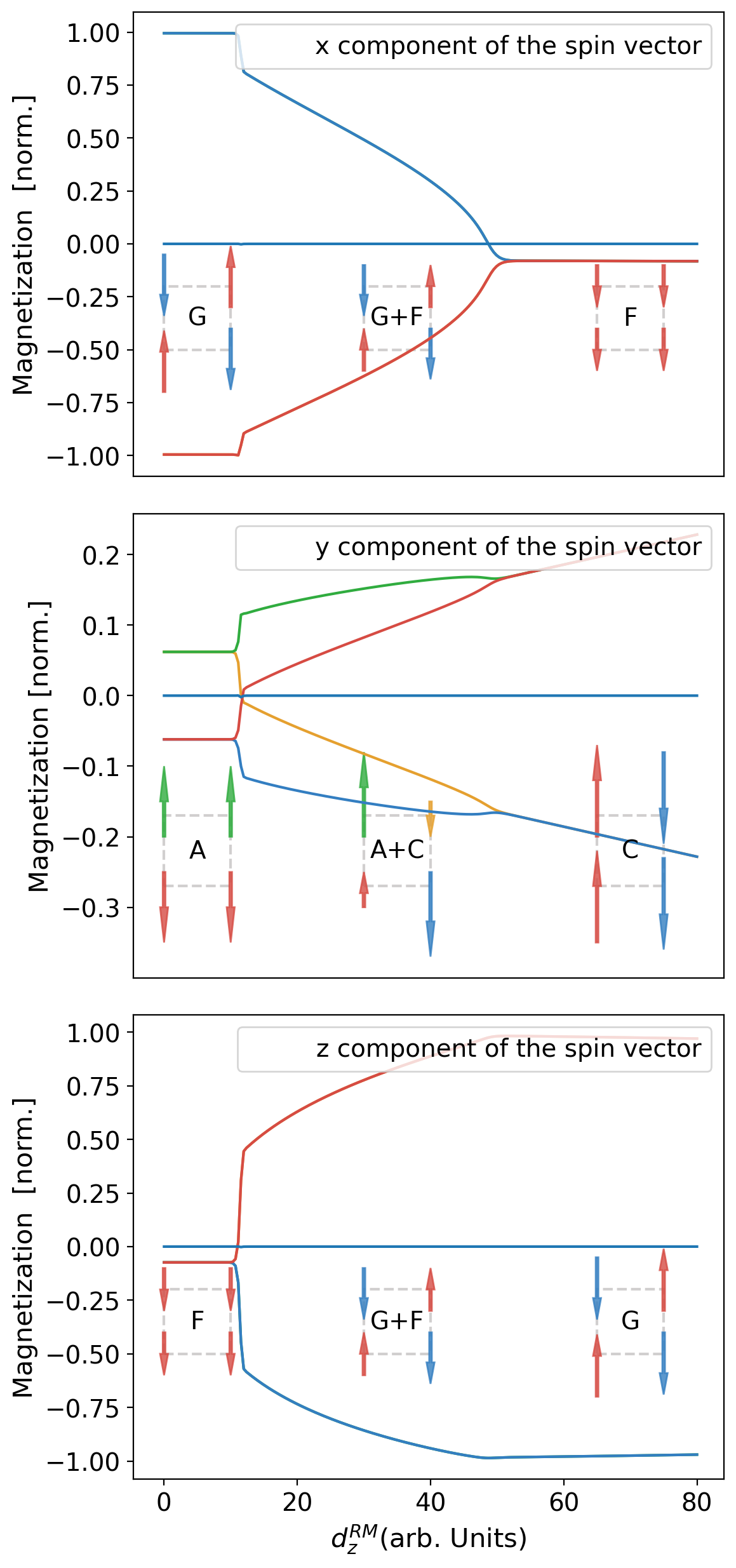} 
	\caption{Plot of the evolution of the $x$, $y$ and $z$ projections of the local $M$ cation spin components versus the DMI strength between $R$ and $M$ cations. A schematic representation of how the $M$ spins look like is also given for the three main phases $\Gamma_4$, $\Gamma_{24}$ and $\Gamma_2$.
	}
	\label{fig:SR_spin}
\end{figure}

To further understand the SR, we have studied how the stability of the magnetic orders is affected by the value of the DMI coupling between $R$ and $M$.
This allows us to determine how the strength of the interaction between $R$ and $M$ spins influences the SR.
Fig.~\ref{fig:Fe_SR} shows the equilibrium state of the structure which is projected to the different irreducible magnetic orders along $x$, $y$ and $z$ directions versus $d_z^{RM}$. (The figure presents three components of the spin as projected to different irreducible representations.)

For values of $d_z^{RM} < 16$ (arb. units) we can see that we have the $\Gamma_4$ state with the main direction of the spin along $x$ with G-type AFM order (G$_x$) and small components (canting) of the spins along $y$ and $z$ directions with A-AFM (A$_y$) and FM (F$_z$) ordering respectively. 
For $43 \text{(arb. units)} > d_z^{RM} > 16$ (arb. units) we have a coexisting region that we denote $\Gamma_{24}$, where mostly $\Gamma_2$ and $\Gamma_4$ states are present. 
The system enters to this state through a sudden jump in magnetic order (we also have a discontinuity in the energy of the system).
As we move towards higher values of $d_z^{RM}$ the $\Gamma_4$ contribution is reduced while $\Gamma_2$ contribution increases up to $d_z^{RM}>43$ (arb. units) where only the $\Gamma_2$ is present, the SR being completed. The transition from $\Gamma_{2}$+$\Gamma_{4}$ to $\Gamma_{2}$ at $d_z^{RM}=43$(arb. units) is continuous.

\begin{figure}[htb!]
	\centering
	\includegraphics[width=1\linewidth ,keepaspectratio=true]{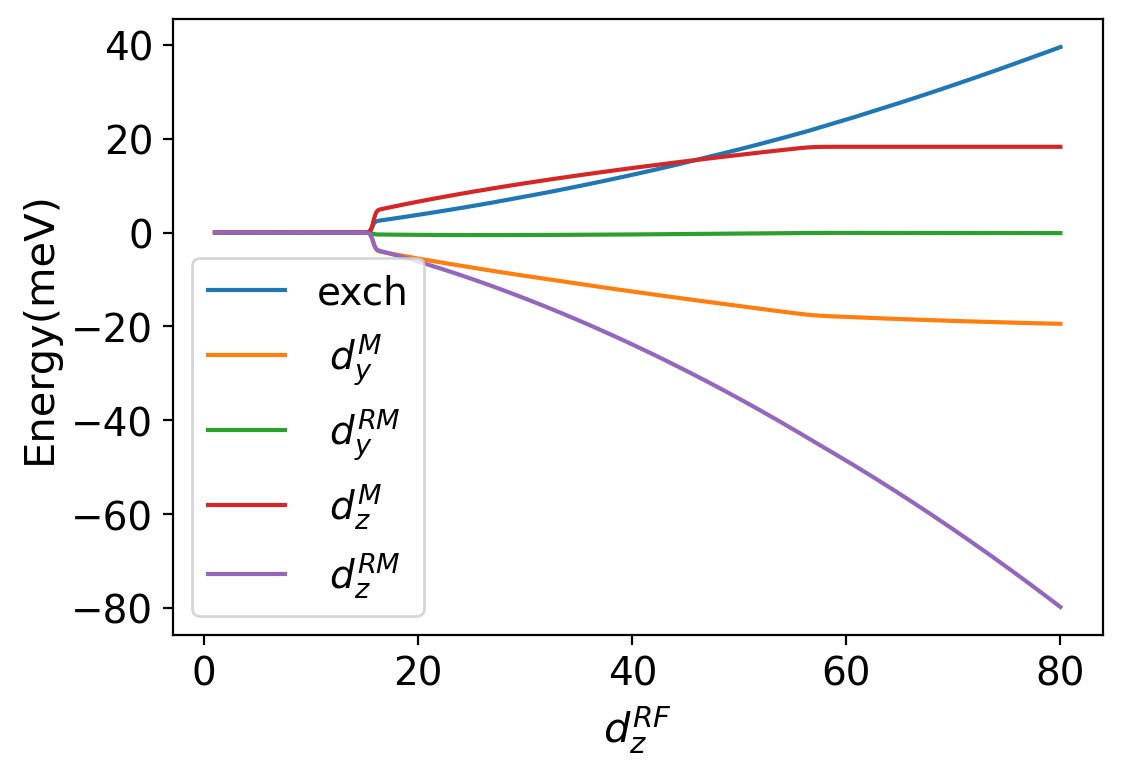} 
	\caption{Decomposition of the total energy from spin dynamics to its components i.e energy from exchange interaction (exch), energy from $y$ componenet of the DMI between M and RM in (d$_y^M$,d$_y^{RM}$) and energy from $z$ componenet of the DMI between M and RM (d$_z^M$,d$_z^{RM}$) . 
	}
	\label{fig:enrg}
\end{figure}

To get further insight into this transition, in Fig.~\ref{fig:SR_spin} we show the evolution of the atomic site projection of the spins in $x$, $y$ and $z$ components.
Since the SR transition is due to the $ d_{z}^{RM}F_{R,x}C_{M,y}$ term in the Hamiltonian, we can observe an increase of the $C$-type canted order along the $y$ direction as the interaction between $R$ and $M$ becomes stronger.
Additionally, since the magnitude of the canted ferromagnetic order on $M$ is constant (FM order along $x$ and $z$ direction before and after SR), this increase in $C$-type order can only come from a reduction of the $G$-type order component of the spin.

\begin{figure}[htb!]
	\centering
	\includegraphics[width=1\linewidth ,keepaspectratio=true]{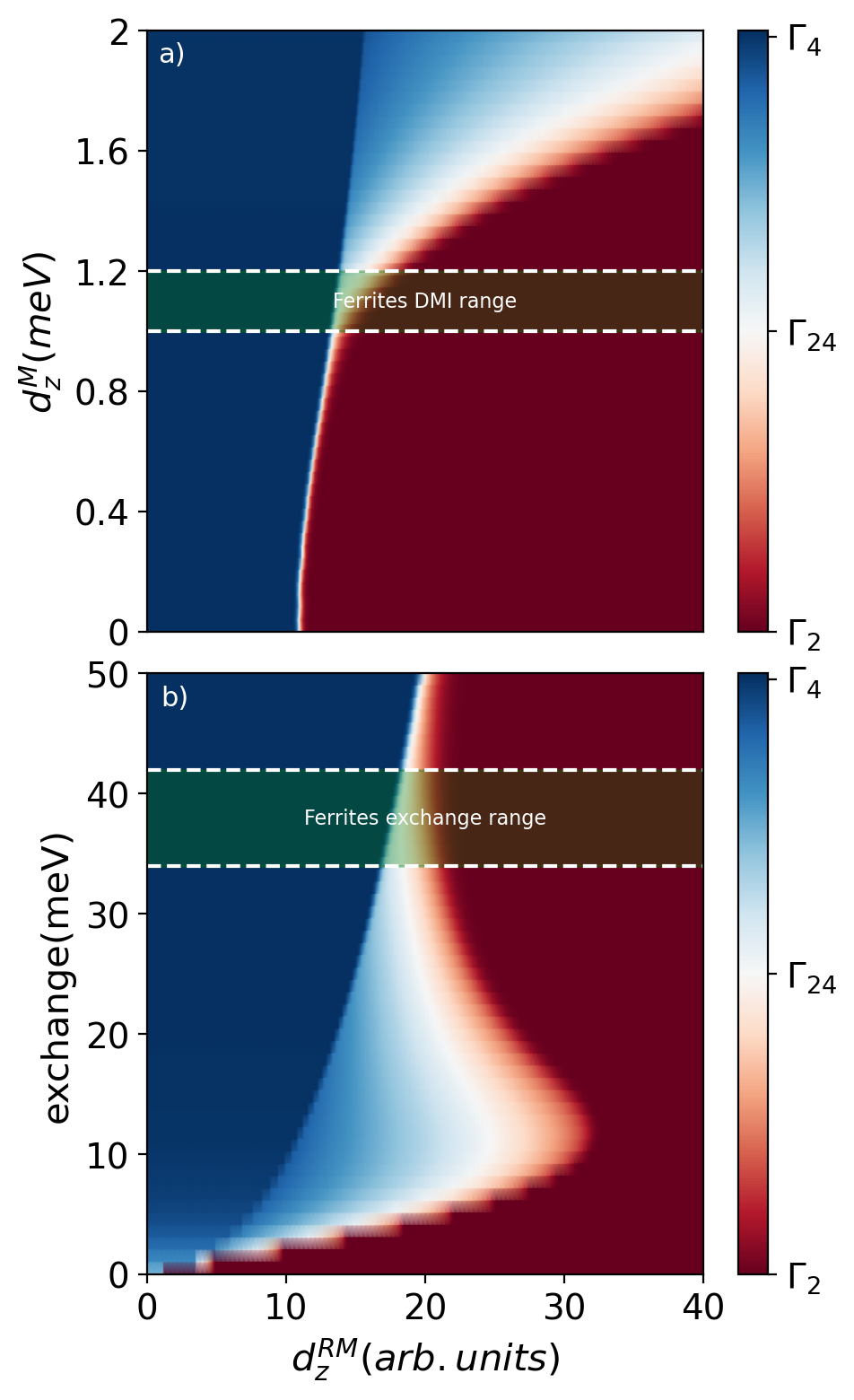} 
	\caption{Phase diagrams for $\Gamma_4$ to $\Gamma_2$ SR as a function of $d_z^{RM}$ (panel a) and $d_z^{M}$ with constant exchange of 38 meV (corresponding to the ferrites) and between $d_z^{RM}$ and exchange interaction of the transition metals (panel b) with constant value of 1.1 meV for  $d_z^{RM}$ interaction. The dashed lines with green background showing the DMI (panel a) and exchange (panel b) of Iron for the whole range of La familly (La to Lu) in RFeO$_3$)
	}
	\label{fig:phase_2}
\end{figure}

In Fig.~\ref{fig:enrg} we show how the different energy contributions of the system (superexchange and DMI) evolve with respect to the $d_{z}^{RM}$ parameter. 
We can see that the contributions coming from $d_z^{M}$ and superexchange interactions between transition metals are positive and increase as we go from $\Gamma_{4}$ to $\Gamma_{2}$ state, which means that they are against the SR. 
In fact these interactions are determinant for how fast the SR happens. 
The exchange interaction is the main interaction that resists against the SR and this is due to the fact that the SR involves an increase of the $C$-type ordering on $M$ sites (via $d_{z}^{RM}F_{R,x}C_{M,y}$, as mentioned above) and the reduction of the $G$-type order, which costs some energy.
Therefore, to overcome this energy penalty we need larger interaction between $R$ and $M$ to complete the SR which is provided by more ordering of the $R$ site atoms.

\subsubsection{Parameters affecting the $\Gamma_4$ to $\Gamma_2$ SR}
\begin{figure}[htb!]
	\centering
	\includegraphics[width=1\linewidth ,keepaspectratio=true]{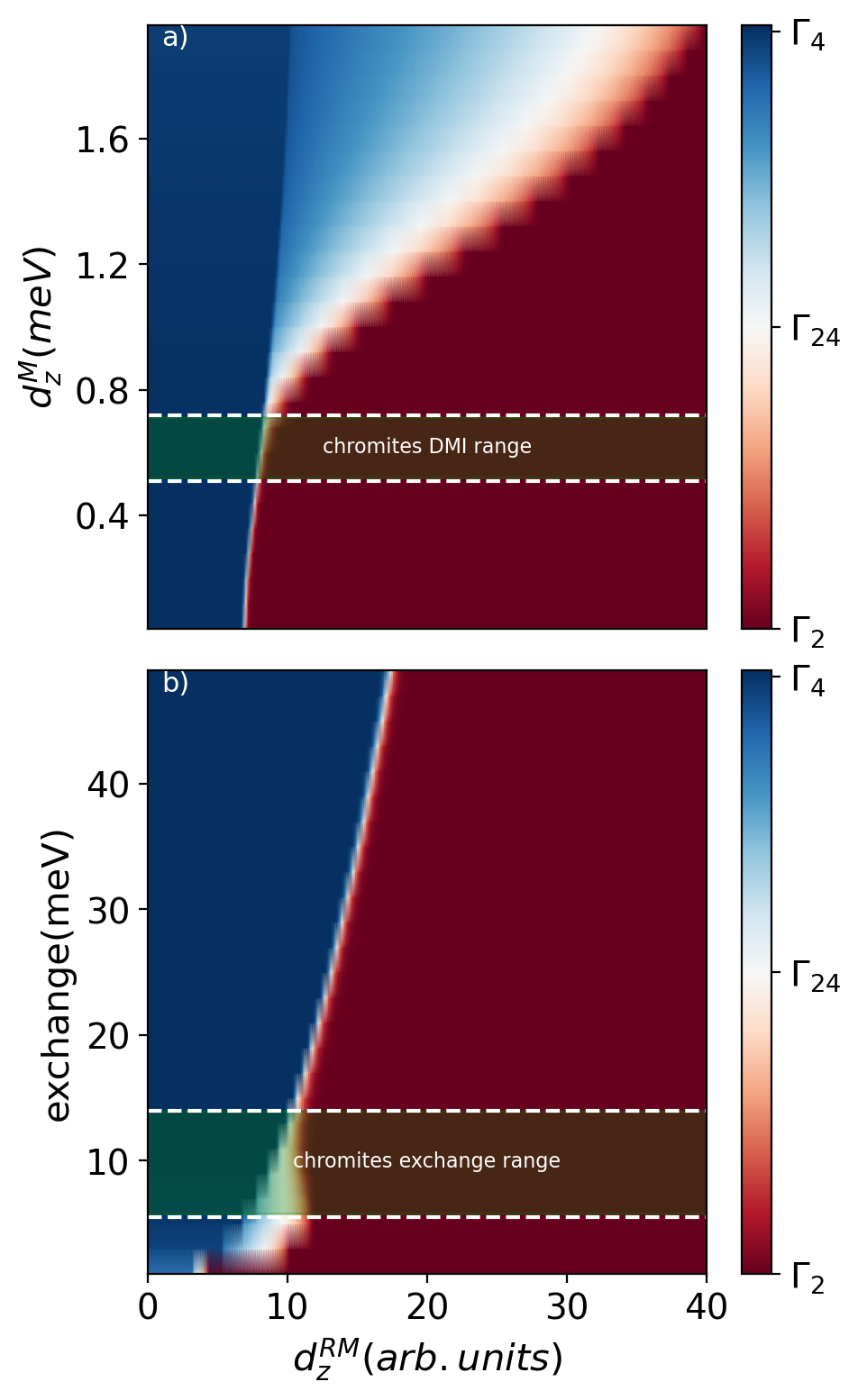} 
	\caption{Phase diagrams for $\Gamma_4$ to $\Gamma_2$ SR as a function of $d_z^{RM}$ (panel a) and $d_z^{M}$ with constant exchange of 9 meV (coresponding to Cr) and between $d_z^{RM}$ and exchange interaction of the transition metals (panel b) with constant value of 0.6 meV for  $d_z^{RM}$ interaction . The dashed lines with green background showing the DMI (panel a) and exchange (panel b) of Cr for the whole range of La familly (La to Lu) in RCrO$_3$)
	}
	\label{fig:phase_2Cr}
\end{figure}

One of the properties that is important to understand is the temperature range in which the spins start and complete their reorientation. 
From our model, we found that three parameters affect how fast the SR happens: the DMI $d_z^{RM}$ between $R$ and $M$ cations (related to the ordering amplitude of the $R$ sites), the DMI   $d_z^{M}$ between $M$ cations, and the superexchange interaction $J^M$ between $M$ cations. 
The ratios between these three parameters drive and determine the energy difference between the $\Gamma_4$ and $\Gamma_2$ states and hence the temperature range where the SR takes place.

To highlight these parameter effects we report 
in Fig.~\ref{fig:phase_2} and \ref{fig:phase_2Cr} a 2D plot showing the presence of the $\Gamma_4$, $\Gamma_2$ and $\Gamma_{24}$
regions with respect to $d_z^M$ and $d_z^{RM}$ values at fixed $J^M$ as calculated for GdFeO$_3$.
Fig.~\ref{fig:phase_2Cr} shows the same but for a fixed value of $J^M$ corresponding to the one calculated for GdCrO$_3$. 
As we can see, for too small values of $d_z^M$ the system only experiences an abrupt transition (first order) between $\Gamma_4$ and $\Gamma_2$ without any coexisting region and the ratio between $d_z^M$ and $d_z^{RM}$ at which the transition appears is rather constant.
However, beyond a critical value of $d_z^M$ a $\Gamma_{24}$ coexisting region appears and grows with the amplitude of $d_z^M$. 
This means that, for a given value of $J^M$, if $d_z^M$ is not large enough the system will never experience a slow SR.
Once the coexistence region opens, it grows very fast with $d^M$ such that for large enough $d_z^M$ and $d_z^{RM}$ values a slow SR transition is always guaranteed.
On the other side, in Fig.~\ref{fig:phase_2}(b) and \ref{fig:phase_2Cr} we can see how the coexisting region area is affected by the value of $J^M$ at a fixed value of $d^M_z$. 
Here we can remark that if $J^M$ is too large or too small then the $\Gamma_{24}$ area is strongly reduced.

In Fig.~\ref{fig:phase_2} and \ref{fig:phase_2Cr} we also draw the maximum and minimum values of $d^M_z$ and $J^M$ as obtained for $R$FeO$_3$ and $R$CrO$_3$, respectively, for the whole series of Lanthanides $R=$La to Lu (horizontal dashed lines). 
We can see that the range of these parameters is not too large and that they cross small areas of the coexistence region where SR is possible.
We can remark that the SR area for $M=$~Cr is particularly small while it is potentially larger for Fe.

These phase diagrams help if one wants to design engineering of SR speed in these crystals. 
For example, if a slower SR is desired, the Fe case will be more interesting through doping with atoms that will reduce the superexchange interactions between irons and/or that will increase the DMI between irons (increase of the wFM).

\subsubsection{Origin of the slow $\Gamma_4$ to $\Gamma_2$ rotation}

\begin{figure}[htb!]
	\centering
	\includegraphics[width=1.1\linewidth ,keepaspectratio=true]{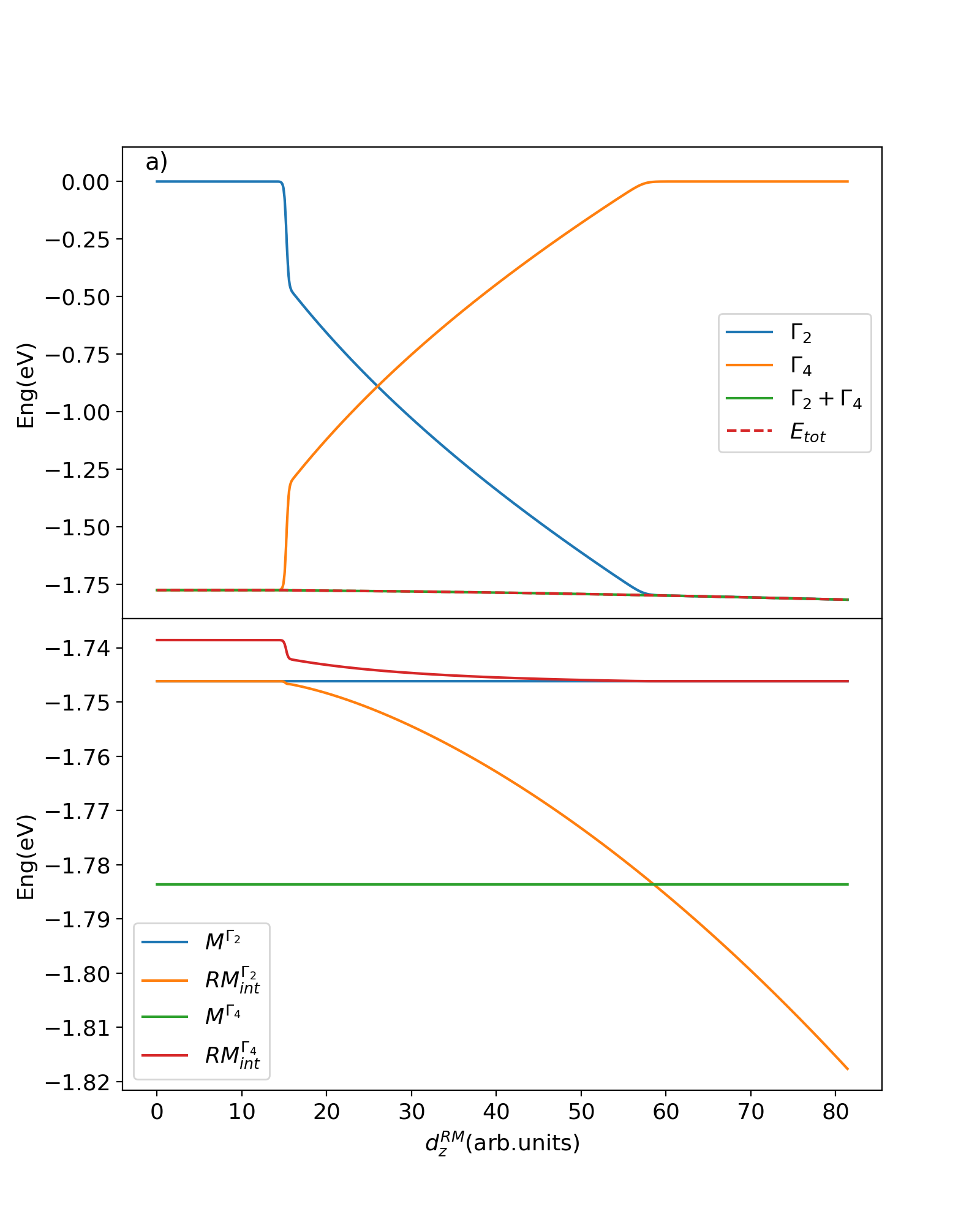} 
	\caption{Decomposition of the energy of the $\Gamma_2$ and $\Gamma_4$ as a function of $d_z^{RM}$. Panel a) shows the energy of $\Gamma_2$ and $\Gamma_4$ and $\Gamma_{2}+\Gamma_4$ which is the sum of the energies of the two states and $E_{tot}$  which is the totla energy from simulations. In panel b), the energy of each state is decomposed into its pure $M$ sublattice contributions and the interaction contribution between $R$ and $M$, the pure $R$-$R$ interactions being neglected. In panel b), the zero energy reference of $M$ and $R$ interactions are taken to be the one of $M^\Gamma_2$. We can notice that $RM$ interacting term is the one that lowers the energy of $\Gamma_2$ by becoming larger than the energy difference between pure $M^\Gamma_4$ and $M^\Gamma_2$}
	\label{fig:eng_decom}
\end{figure}

\begin{figure}[htb!]
	\centering
	\includegraphics[width=1\linewidth ,keepaspectratio=true]{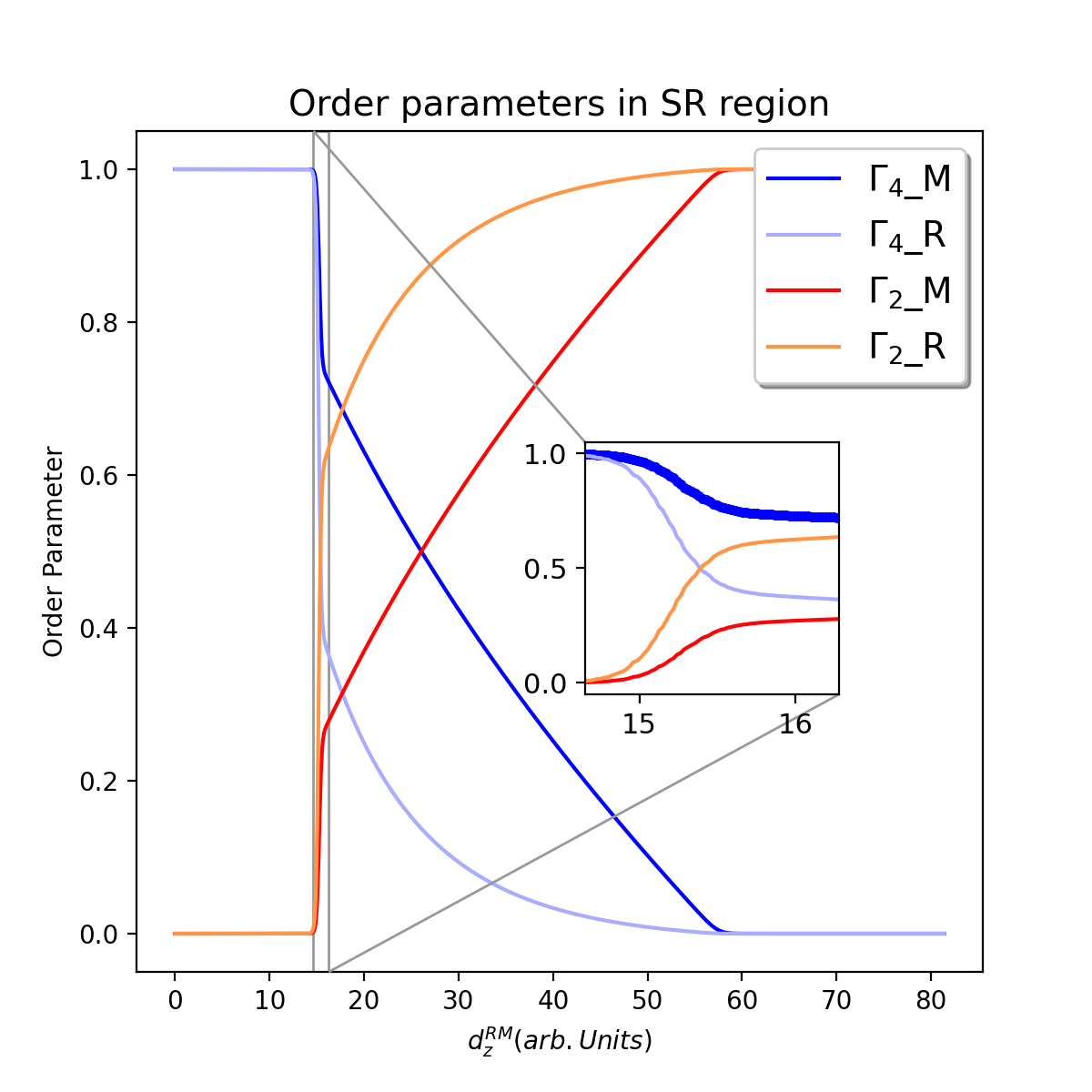} 
	\caption{Order parameter ($\Gamma_2$ with orange color and $\Gamma_4$ with blue color) decomposition into $M$ and $R$ sublattices when crossing a SR region (here as a function of $d_z^{RM}$). The inset shows the area where the SR starts and we can notice the change of order of R site first that drags the M site order afterward.
	}
	\label{fig:ordr_param}
\end{figure}

As we show in the analytical part of our model, there is no interaction between $\Gamma_2 $ and $\Gamma_4$, and we have $<\Gamma_2|H|\Gamma_4>=0$.
This would mean that the transition should be fast from our model since, without interaction between the two states, there is apparently no reason why their coexistence will reduce the energy, and we should have an sharp transition between them. However, our simulations based on this non-interacting Hamiltonian show that a coexisting region exists where both $\Gamma_2$ and $\Gamma_4$ are present together. 

To figure out what is happening, we plot in Fig.~\ref{fig:eng_decom}(a) the energy change with respect to the $d_z^{RM}$ as decomposed into a pure $\Gamma_4$, pure $\Gamma_2$ , the sum of the energy of $\Gamma_2$ and $\Gamma_4$ and total energy from our simulations E$_{tot}$. To understand if there is hidden coupling between $\Gamma_2$ and $\Gamma_4$ states, we have plotted the sum of energies of $\Gamma_2$ and $\Gamma_4$ ($\Gamma_2+\Gamma_4$) and total energy from our simulations E$_{tot}$ and as we can see, the two energies match exactly which proves that although there is no coupling between the two states, the SR is slow, . 

In Fig.~\ref{fig:eng_decom}(b) we report the energy decomposition to $M$ sublattice only ($M^{\Gamma_4}$, $M^{\Gamma_2}$) and interaction between R and M in each state (${RM}^{\Gamma_4}_{int}$, ${RM}^{\Gamma_2}_{int}$,).  We can see that the $M$ sublattice energy of the $\Gamma_2$ state ($M^{\Gamma_2}$, blue line) is higher than the energy of the $M$ sublattice in the $\Gamma_4$ state ($M^{\Gamma_4}$, green line), as expected since the $\Gamma_4$ phase is the ground state when only the $M$ sublattice is considered. 
The two $RM_{int}^{\Gamma_4}$ (red line) and $RM_{int}^{\Gamma_2}$ (orange line) interaction terms clearly show that $RM_{int}^{\Gamma_2}$ lowers the energy of the $\Gamma_2$ phase with an amplitude that can compensate the energy difference between $M^{\Gamma_4}$ and $M^{\Gamma_2}$, such that the $\Gamma_2$ phase can be lower in energy than the $\Gamma_4$ phase.
This also proves that the $M$ sublattice alone prefers to stay in the $\Gamma_4$ state while the $R$ sublattice pushes the $M$ sublattice to be in the $\Gamma_2$ state (the $RM_{int}^{\Gamma_2}$ energy is stronger and more negative than the $RM_{int}^{\Gamma_4}$ energy).

In Fig.~\ref{fig:ordr_param} we show the evolution of the order parameters but decomposed into sublattice contributions ($\Gamma_4$\_M, $\Gamma_4$\_R , $\Gamma_2$\_M and $\Gamma_2$\_R). 
We can observe that in the SR region, when going from $\Gamma_4$ to $\Gamma_2$ the $R$ spins start to rotate first and they drag the $M$ sublattice afterward (highlighted in the inset).
Since the $M$ sublattice prefers to stay in the  $\Gamma_4$ state while the $R-M$ interaction favors the $\Gamma_2$ state, the system ends up in a mixed state even if no $\Gamma_{2}-\Gamma_{4}$ interaction is present in the Hamiltonian. 
To understand this better, note that in this problem we do no really have two competing orders ($\Gamma_{2}$ and $\Gamma_{4}$), but four (($\Gamma_4(M)$, $\Gamma_4(R)$, $\Gamma_2(M)$ and $\Gamma_2(R)$). As we vary the key Hamiltonian parameter in Fig.~\ref{fig:ordr_param}, the $\Gamma_2(R)$ order becomes favorable over $\Gamma_{4}(R)$; we thus have a $\Gamma_{4}\rightarrow\Gamma_{2}$ rotation of the $R$ sublattice (accompanied by a relatively tiny $\Gamma_{2}(M)$ component) that yields a reduction of the energy as compared to a pure $\Gamma_{4}$ state. Eventually, the $\Gamma_2(R)$ order grows and drags the $M$ spins to rotate as well, the final result being a pure $\Gamma_{2}$ state.

\subsubsection{Effect of SIA on $\Gamma_4$ to $\Gamma_2$ SR}

\begin{figure}[htb!]
	\centering
	\includegraphics[width=1\linewidth ,keepaspectratio=true]{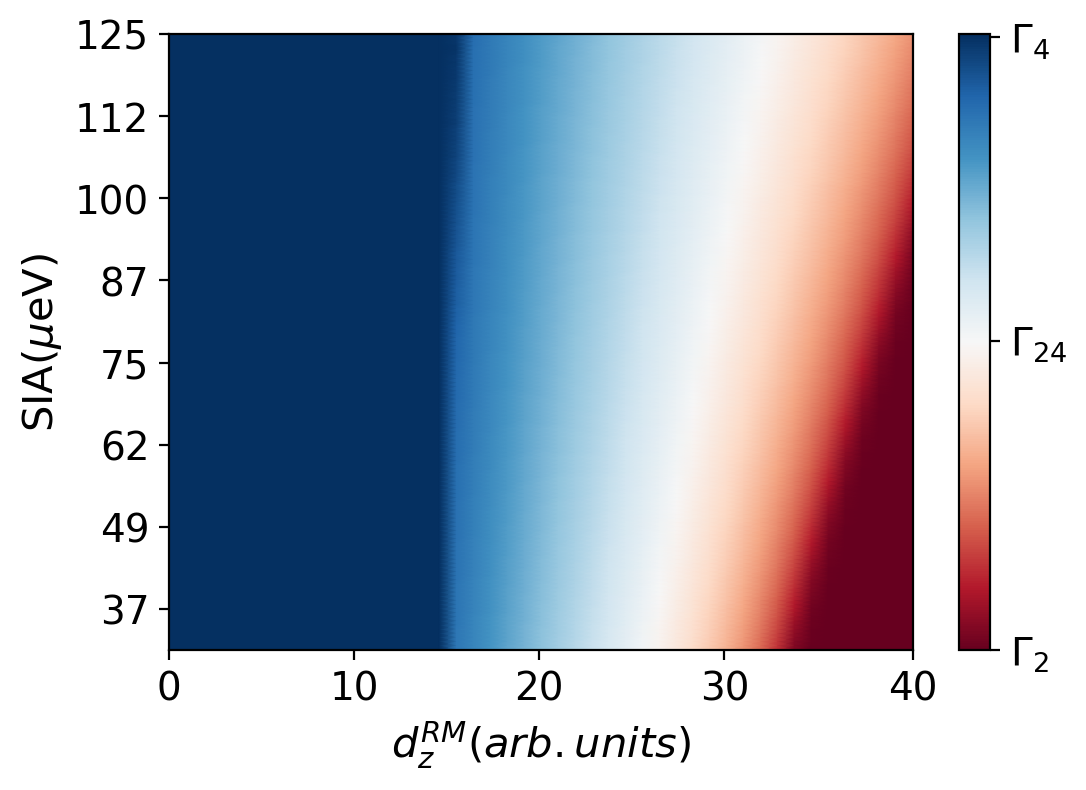} 
	\caption{Effect of SIA on the $\Gamma_4/\Gamma_2$ SR phase diagram. Horizontal axis shows the amplitude of $z$ component of the DMI between $R$ and $M$ sites while the vertical axis shows the amplitude of the SIA of the $M$ site. 
	}
	\label{fig:SIA}
\end{figure}

Although the SIA amplitude on the $M$ site is not very large, we can probe it from our model and have an estimate of its effect on SR.
To that end, we report in Fig.~\ref{fig:SIA}(a) the phase diagram of the $\Gamma_4$, $\Gamma_2$ and $\Gamma_{24}$ presence with respect to $d^{RM}$ and SIA of $M$. 
We can see on this plot that the SIA does not change the $\Gamma_4$ to $\Gamma_{24}$ transition position; in contrast, when the SIA increases, it has the tendency to increase the $\Gamma_{24}$ SR area at the expense of the $\Gamma_{2}$ state. However, the effect of the SIA is much smaller than the ones of $J^M$, $d^M$ or $d^{RM}$.

 \subsubsection{Summary for the $\Gamma_{4}$ to $\Gamma_{2}$ transition}

In summary, in this section we have shown that our model well reproduces the temperature dependent SR. This behaviour shows that the SR is directly linked to the ordering of the rare earth in ferromagnetic order and proves that the mechanism behind the SR in $RM$O$_3$ compounds is not related to the SIA \cite{white-1969}, but it is the DMI between $R$ and $M$ that drives this SR.

A study by Vibhakar {\it et al}. \cite{Vibhakar-2020} on triple A-site columnar-ordered quadruple perovskites has shown that the mechanism behind SR in these structures is the competition between DMI and SIA,  which is similar to the mechanism that we found to be at play in $RM$O$_3$'s SR.

In our simulations, we have also studied how different parameters affect the speed of SR.
We can say that the presence of a smooth transition between $\Gamma_4$ and $\Gamma_2$ phases through a coexisting region $\Gamma_{24}$ is very subtle and depends on the ratio between $J^M$, $d^M_z$ and $d^{RM}_z$ interactions. If $d^{M}_z$ is zero, a transition between $\Gamma_4$ and $\Gamma_2$ can exist but only through a first order abrupt change; the $d^{M}_z$ interaction is mandatory to have a smooth SR transition.

\subsection{ $\Gamma_4$ to $\Gamma_1$ reorientation:}

\begin{figure*}[htb!]
	\centering
	\includegraphics[width=\textwidth ,keepaspectratio=true]{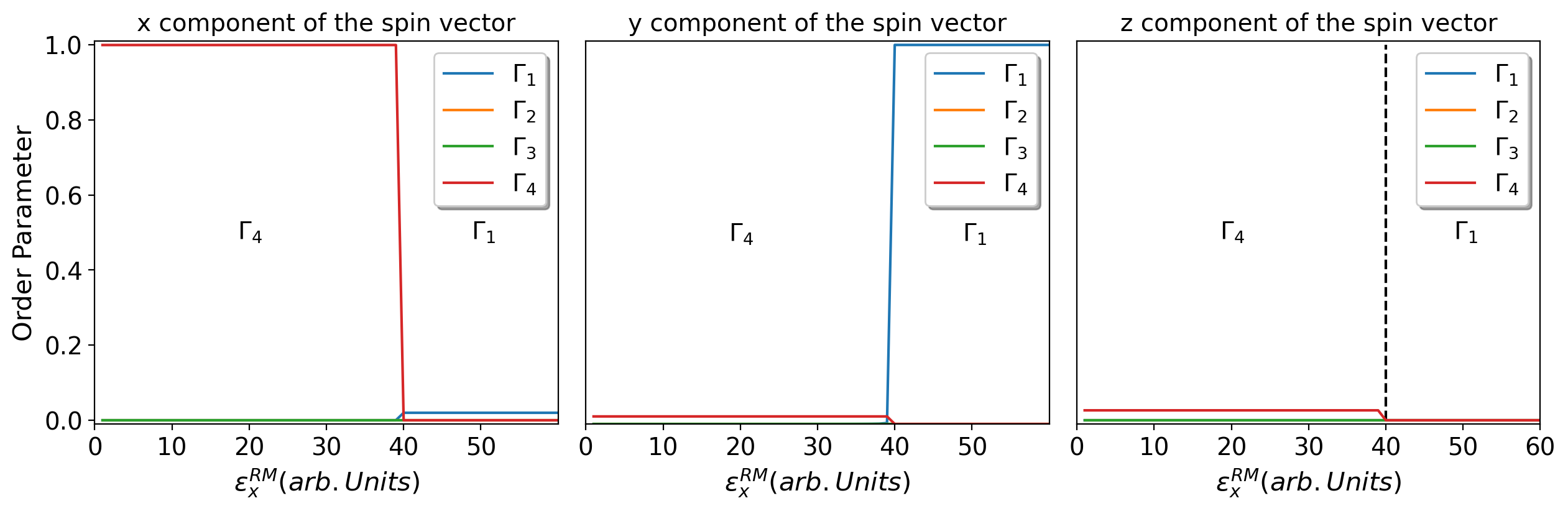} 
	\caption{Magnetic structure for $\Gamma_4$ to $\Gamma_1$ SR decomposed to different irreducible representations for different components of the spin in $x$, $y$ and $z$ directions. The horizontal axis showing the magnitude of ASE in x direction between $R$ and $M$ and vertical axis showing the order parameter magnitude normalised.
	}
	\label{fig:G4_G1}
\end{figure*}
To explain the $\Gamma_4$ to $\Gamma_1$ SR we need a strong interaction between the $R$ and $M$ sites within the $\Gamma_1$ state to allow the $M$ site order to go from its energetically favourable state $\Gamma_4$ to the less energetically favourable state $\Gamma_1$.
If not, we would have each sublattice ordering in different direction like what is observed experimentally in TbFeO$_3$ \cite{Cao-2016}. 
However, according to our model, the sole interactions between $R$ and $M$ atoms in the $\Gamma_1$ state are $-8Nd_{x}^{RM}C_{R,z}\bar{G}_{M,y}$ and $-8Nd_{y}^{RM}C_{R,z}A_{M,x}$. 
The second term is the coupling between $A_{M,x}$ and $C_{R,z}$, which is small since the $A_{M,x}$ canting is very small compared to $\bar G_{M,y}$.
Hence, the only remaining term which can make this SR possible is $-8Nd_{x}^{RM}C_{R,z}\bar{G}_{M,y}$. 
From our DFT calculations and from symmetry analysis (since these parameters are originating from  $\overset{-}{a}\overset{-}{a}\overset{+}{c}$ oxygen octahedra rotations as discussed in previous section) we know that $d_x^{RM}$ is very small (see table \ref{tab:DMI}) such that it is not possible to explain the $\Gamma_4$ to $\Gamma_1$ SR using this interaction.

So far we have neglected the anisotropic spin exchange interactions (ASE) in our model  because the effects from these interactions are often negligible with respect to the superexchange or DMI. 
Now that we have the DMI small too, we will consider the ASE to check whether it can take some importance while the DMI is small. The definition of DMIs and ASE vector components are as follows:
\begin{equation} \label{eq:17}
\begin{array}{l@{}l}
d_x^{ab}=\frac{1}{2}(J_{yz}^{ab}-J_{zy}^{zb})
\end{array}
\end{equation}
\begin{equation} \label{eq:18}
\begin{array}{l@{}l}
\varepsilon_x^{ab}=\frac{1}{2}(J_{yz}^{ab}+J_{zy}^{ab}),
\end{array}
\end{equation}
where $\varepsilon_x^{ab}$($d_x^{ab}$) represents the ASE (DMI) vector component in $x$ direction between atom $a$ and atom $b$ and $J_{yz}$ is the exchange interaction between spins directing in $y$ direction on atom $a$ and in $z$ direction on atom $b$ (other component, i.e in $y$ and $z$ directions, can be obtained by cyclic permutation of the $xyz$ directions). 
We can see that when a component of the DMI vector is small it is  probable to have the ASE vector for that component to be bigger depending on the magnitude of  $J_{yz}$.

In Table \ref{tab:ASE} we show the calculated ASE between Fe sites and Cr sites. The calculated results show that  the $x$ component of the ASE vector is the largest with respect to the $y$ and $z$ components for Fe-Fe and Cr-Cr atom pairs.
In Table \ref{tab:ASE} we also report the calculated ASE vector for Fe-Gd and Cr-Gd pairs. 
The biggest component of the ASE vector is in the $y$ direction for Gd-Fe while it is along the $x$ direction for the Gd-Cr case.
We note that the ASE interaction in the Hamiltonian takes the same place as the DMI does, i.e for the $\Gamma_1$ state we have:
\begin{equation} \label{eq:19}
\begin{array}{l@{}l}
-8N\varepsilon_{x}^{RM}C_{R,z}\bar{G}_{M,y} 
\end{array}
\end{equation}
\begin{equation} \label{eq:20}
\begin{array}{l@{}l}
8N\varepsilon_{y}^{RM}C_{R,z}A_{M,x}
\end{array}
\end{equation}
Considering these ASE interactions, we can say that the $\Gamma_4$ to $\Gamma_1$ SR can happen through the $x$ component of the ASE (through $-8N\varepsilon_{x}^{RM}C_{R,z}\bar{G}_{M,y}$  interaction), which will take the place of the DMI when the latter is small.
This conclusion is in agreement with Zvezdin~\cite{zvezdin-1979} who explained the origin of the $\Gamma_4$ to $\Gamma_1$ SR to originate from ASE.

\begin{table}
\caption{Calculated ASE components from DFT of GdFeO$_3$ and GdCrO$_3$ (units are meV). $\varepsilon_i$ are the ASE vector components along $i=$ $x$, $y$ and $z$.} 
\label{tab:ASE}
\centering
 \begin{tabularx}{\columnwidth}{ >{\setlength\hsize{1\hsize}\centering}X >{\setlength\hsize{1\hsize}\centering}X >{\setlength\hsize{1\hsize}\centering}X c}
\hline
\hline
     & $\varepsilon_x$ & $\varepsilon_y$ & $\varepsilon_z$ \\
\hline \xrowht{6pt}
Fe-Fe & -0.015  & 0.000 & 0.000    \\
 \xrowht{6pt}
Cr-Cr & 0.051   & 0.000 & 0.003   \\
 \xrowht{6pt}
Gd-Fe & 0.006  & -0.016 & 0.007   \\
 \xrowht{6pt}
Gd-Cr & -0.030 &  0.010 & -0.009  \\
\hline
\end{tabularx}
\end{table}

\begin{figure}[htb!]
	\centering
	\includegraphics[width=1\linewidth ,keepaspectratio=true]{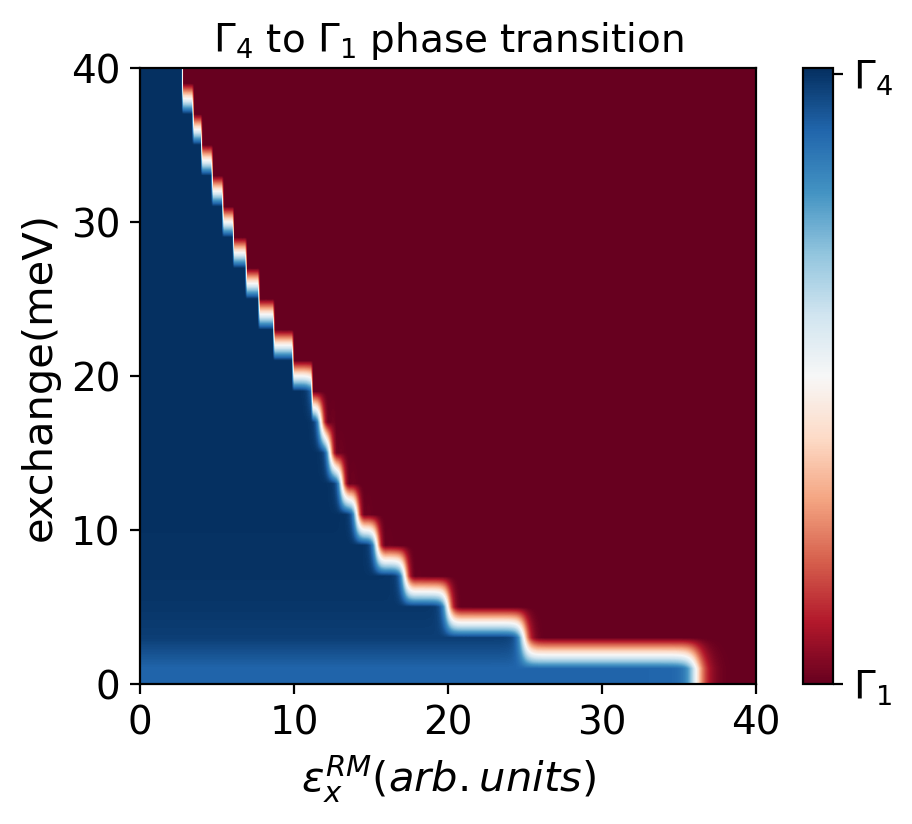} 
	\caption{Phase diagrams of $\Gamma_4$ to $\Gamma_1$ transition by plotting exchange on transition metal sites (J$^M$) vs the anisotropic exchange $\varepsilon_x^{RM}$. We can see that the SR transition is abrupt for the whole range of exchange.}
	\label{fig:G41_Phase}
\end{figure}

We will now study this SR by tuning the $\varepsilon_x^{RM}$ ASE coupling in our model. In Fig.~\ref{fig:G4_G1} and \ref{fig:G41_Phase} we report how the relative $\Gamma_4$ and $\Gamma_1$ stability evolves with respect to the $\varepsilon_x^{RM}$ parameter at 0~K (ground state). 
In contrast to the $\Gamma_4$ to $\Gamma_2$ transition we can see that there is no coexisting region between $\Gamma_4$ and $\Gamma_1$ states, the transition is always abrupt with respect to the $\varepsilon_x^{RM}$ amplitude.
To confirm this, we also explore in Fig.~\ref{fig:G4_G1} how the superexchange parameter $J^M$ affects the $\Gamma_4$ to $\Gamma_1$ SR transition.
We can clearly see that whatever value of $J^M$ we considered, the $\Gamma_4$ to $\Gamma_1$ SR is always abrupt without any coexisting region.
We can also remark that $J^M$ favours the $\Gamma_1$ state with respect to the $\Gamma_4$ state, which can be logically understood by the fact that in the $\Gamma_1$ phase all the directions are AFM and, since the torque creating this SR is acting on $G_{M,y}$ (in contrast to $\Gamma_4$ to $\Gamma_2$ SR where the torque is acting on $C_{M,y}$), the cantings are smaller than in the $\Gamma_4$ state.
Hence, unlike the $\Gamma_4$ to $\Gamma_2$ case, the SR involving $\Gamma_{1}$ happens as soon as the system overcomes the energy difference due to ASE and SIA between the two states, making this transition abrupt.

We also need to mention that this mechanism explains the Ising-like (strongly colinear) nature of the $\Gamma_1$ state \cite{zvezdin-1979}. 
Since the force creating this SR is acting between $G$-type order of $M$ site and $C$-type order of $R$ site (i.e $-8N\varepsilon_{x}^{RM}C_{R,z}\bar{G}_{M,y}$), the $M$ atoms in $\Gamma_1$ state will have a very small canting compared to other states and the spins will mainly order in the G$_y$ type, hence closer to an Ising like nature. 

\section{Conclusion}

We have studied in this paper the microscopic mechanism behind the SR and MR magnetic behaviours of the $RM$O$_3$'s through a Heisenberg model where we considered the superexchange interactions and DMI between the transition-metal sites, as well as between the rare-earth ($R$) and transition-metal sites ($M$), and we neglected the superexchange and the DMI between the $R$ spins as they are much smaller than the other interaction parameters.

 We conclude that there are two interactions polarising the $R$ atom site, i.e (i) the superexchange between $M$ sites (through its wFM) and $R$ sites and (ii) the DMI between $R$ and $M$, which can results into two effects. 
Indeed, we can have that both interactions polarise the $R$ element parallel to the $M$ wFM canting direction such that there will be no MR but an amplification of the total magnetization of the crystal (See Fig.~\ref{fig:SR-Mr} panel b blue curve). 
We can also have that both interactions polarise the $R$ element in opposite direction to the wFM of the $M$ cation such that the total magnetization amplitude can be reduced up to a critical temperature below which its sign change (See Fig.~\ref{fig:SR-Mr} panel b red curve). The change of sign appears when the negative $R$ cation magnetization compensates the positive one of the $M$ site (wFM).

Our analysis of the SR transitions have shown that the $\Gamma_4$ to $\Gamma_2$ transition similarly comes mainly from the DMI interactions between $M$ site and $R$ site but it can be weighted by the superexchange between the $M$ sites. 
We found that within a relatively wide range of these three interactions this SR transition is smooth and happens through a mixed state where the $\Gamma_4$ and the $\Gamma_2$ phases coexist even though they do not interact in our Hamiltonian.
How broad is the temperature range in which the SR takes place through the $\Gamma_{24}$ mixed state depends on a subtle ratio between DMI and exchange interactions between $M$ sites, which can vary depending on the rare earth and transition metal cation that is present in the $Pnma$ perovskite structure.
We also found that the $\Gamma_4$ to $\Gamma_1$ SR transition depends on even more subtle interactions (anisotropic superexchange that acts as the DMI) but, contrary to the $\Gamma_4$ to $\Gamma_2$ SR, it never presents a coexisting region, i.e. it always proceeds through an abrupt change.

The model we have presented can help in designing the strength and amplitude of SR and MR in $RM$O$_3$ through, e.g. doping, strain or pressure that would tune the ratio between the key interactions as desired.
Our model can also be easily extended by including the interactions between the rare-earth spins to study the complex magnetic phase diagrams below the N\'eel temperature of the rare-earth sublattice.
It can be enlarged too with the anisotropic exchanges (important for the $\Gamma_4$ to $\Gamma_1$ SR) \cite{zvezdin-1979} or with a 4-spin interaction term, which has been shown to be important in rare-earth manganites \cite{fedorova2015}.
Because it contains all the key interactions that allow to describe most of the important magnetic properties of $RM$O$_3$ compounds, the model can be used to study dynamically magnetic domain walls. 
Going beyond, the model can be coupled with a lattice model (second principles \cite{wojdel2013, wojdel2016}) to have access to a full atom plus spin dynamics for the simulations of, e.g., recent ultrafast laser excitation experiments made on these crystals \cite{afanasiev2021,tang2018,juraschek2017}.

\section*{Acknowledgements}
The authors thank He Xu for his help in using the TB2J code. 
This work has been funded by the Communaut\'e Fran\c{c}aise de Belgique (ARC AIMED G.A. 15/19-09).
EB and AS thanks the FRS-FNRS for support. J.\'I. thanks the support of the Luxembourg National Research Fund through Grant No. FNR/C18/MS/12705883/REFOX.
The authors acknowledge the CECI supercomputer facilities funded by the F.R.S-FNRS (Grant No. 2.5020.1), the Tier-1 supercomputer of the F\'ed\'eration Wallonie-Bruxelles funded by the Walloon Region (Grant No. 1117545) and the OFFSPRING PRACE project.

\bibliography{Mag.bib}

\end{document}


\title{Magnetic phase diagram of rare-earth orthorhombic perovskite oxides: Supplemental information}

\author{Alireza Sasani}
\affiliation{Physique Th\'eorique des Mat\'eriaux, QMAT, CESAM, Universit\'e de Li\`ege, B-4000 Sart-Tilman, Belgium}

\author{Jorge I\~niguez}
\affiliation{MaterialsResearch and Technology Department, Luxembourg Institute of Science and Technology (LIST), 5 avenue des Hauts-Fourneaux, L-4362, Esch/Alzette, Luxemburg}
\affiliation{Department of Physics and Materials Science, University of Luxembourg, Rue du Brill 41, L-4422 Belvaux, Luxembourg}

\author{Eric Bousquet}
\affiliation{Physique Th\'eorique des Mat\'eriaux, QMAT, CESAM, Universit\'e de Li\`ege, B-4000 Sart-Tilman, Belgium}
\maketitle

To model the magnetic properties of RMO$_3$ perovskite structures, we have used the following Hamiltonian which includes magnetic interactions for transition metal atoms, H$^{MM}$, and magnetic interaction between rare earth atoms H$^{RR}$, and the interaction between the two sublattices H$^{RM}$ (Eq.\ref{eq:1}).  

\begin{equation} \label{eq:1}
\begin{split}
H=H^{MM}+H^{RM}+H^{RR}
\end{split}
\end{equation}

The Hamiltonian for each of the sublattices can be written as following.\\
For the transition metal atoms we would have:
\begin{equation} \label{eq:2}
\begin{split}
H^{MM}=H_{ex}^{MM}+H_{DMI}^{MM}+H_{SIA}^{MM},
\end{split}
\end{equation}
For the $R$ atoms the DMI and exchange interactions are very small and can be neglected such that:
\begin{equation} \label{eq:3}
\begin{split}
H^{RR}\approx H_{SIA}^{RR},
\end{split}
\end{equation}
and for the interaction between two sublattices we have:
\begin{equation} \label{eq:4}
\begin{split}
H^{RM}=H_{ex}^{RM}+H_{DMI}^{RM},
\end{split}
\end{equation}
where the exchange and DMI interactions are considered to have the following forms respectively:

\begin{equation} \label{eq:6}
H_{Ex}^{ab}=\frac{1}{2}\sum_{ij}^N\left( J_{ab,ij}S_{i,a}.S_{j,b}  \right) \\
\end{equation}
\begin{equation} \label{eq:7}
H^{ab}_{dmi}=\frac{1}{2}\sum_{i,j}\left(D_{ab,ij}\times S_{j,a}\right).S_{i,b}\\
\end{equation}
and the SIA is considered to have the following form on both magnetic sublattices:
\begin{equation} \label{eq:8}
H^{aa}_{SIA}=\sum_{i}K_{a}\left(S_{i,a}.\hat{e}_{i}\right)^2,\\
\end{equation}
where i and j are magnetic lattice sites and the some is over magnetic sites with spin  $S$ and the $\hat{e}_i$ shows the SIA direction.

For these structures we have four different symmetry adapted spin states that can be written as shown in Eq.~\ref{eq:9a} to Eq.~\ref{eq:9d}.

\begin{equation} \label{eq:9a}
\begin{array}{l@{}l}
S_{i,a}^{\Gamma_1} &{}=A_{a,x}(-1)^{(n_{z}^{i})}+\bar{G}_{a,y}(-1)^{(n_{x}^{i}+n_{y}^{i}+n_{z}^{i})}+C_{a,z}(-1)^{(n_{x}^{i}+n_{y}^{i})}\\
\end{array}
\end{equation}

\begin{equation} \label{eq:1b}
\begin{array}{l@{}l}
S_{i,a}^{\Gamma_2} &{}=F_{a,x}+C_{a,y}(-1)^{(n_{x}^{i}+n_{z}^{i})}+\bar{G}_{a,z}(-1)^{(n_{x}^{i}+n_{y}^{i}+n_{z}^{i})}\\
\end{array}
\end{equation}

\begin{equation} \label{eq:1c}
\begin{array}{l@{}l}
S_{i,a}^{\Gamma_3}=C_{a,x}(-1)^{(n_{y}^{i}+n_{z}^{i})}+F_{a,y}+A_{a,z}(-1)^{(n_{z}^{i})}\\
\end{array}
\end{equation}

\begin{equation} \label{eq:9d}
\begin{split}
S_{i,a}^{\Gamma_4}=\bar{G}_{a,x}(-1)^{(n_{x}^{i}+n_{y}^{i}+n_{z}^{i})}+A_{a,y}(-1)^{(n_{z}^{i})}+F_{a,z}\\
\end{split}
\end{equation}

Where the $\bar{G}$ shows the main spin direction while the other letters showing small cantings.\\

from now on we will only consider the state with G type order which is the dominant order in RMO$_3$s (i.e $\Gamma_1$, $\Gamma_2$ and $\Gamma_4$).
The general spin state of the atoms in these representations could be written as Follows (Eq.\ref{eq:10}):
\begin{equation} \label{eq:10}
\begin{split}
S_{i,a}=C^1_{\alpha,a}S_{i,a}^{\Gamma_1}+C^2_{\alpha,a}S_{i,a}^{\Gamma_2}+C^4_{\alpha,a}S_{i,a}^{\Gamma_4},
\end{split}
\end{equation}
where $C^g_{\alpha,a} $ is the ratio of the spins in g state.

Using these definitions we can write the exchange interactions for atom $a$ as (Eq.~\ref{eq:11})
\begin{equation} \label{eq:11}
\begin{split}
H^{aa}_{Ex}
&{}=\frac{1}{2}\sum_{i,j}J_{ij}S_{i,a}.S_{j,a}\\
      &{}=\frac{1}{2}\sum_{i,j}\sum_{\alpha}J_{ij}(C^1_{\alpha,a}S_{i,\alpha}^{\Gamma_1}+C^2_{\alpha,a}S_{i,\alpha}^{\Gamma_2}+C^4_{\alpha,a}S_{i,\alpha}^{\Gamma_4}).(C^1_{\alpha,a}S_{i,\alpha}^{\Gamma_1}+C^2_{\alpha ,a}S_{j,\alpha }^{\Gamma_2}+C^4_{\alpha ,a}S_{j,\alpha }^{\Gamma_4})\\
      &{}=\frac{1}{2}\sum_{i,j}\sum_{\alpha}J_{ij}(C^1_{\alpha,a}C^1_{\alpha ,a}S_{i,\alpha}^{\Gamma_1}S_{j,\alpha }^{\Gamma_1} +C^2_{\alpha,a}C^2_{\alpha ,a}S_{i,\alpha}^{\Gamma_2}S_{j,\alpha }^{\Gamma_2} +C^4_{\alpha,a}C^4_{\alpha ,a}S_{i,\alpha}^{\Gamma_4}S_{j,\alpha }^{\Gamma_4}\\ &{}+2C^1_{\alpha,a}C^2_{\alpha ,a}S_{i,\alpha}^{\Gamma_1}S_{j,\alpha }^{\Gamma_2}+2C^1_{\alpha,a}C^4_{\alpha ,a}S_{i,\alpha}^{\Gamma_1}S_{j,\alpha }^{\Gamma_4}+2C^2_{\alpha,a}C^4_{\alpha ,a}S_{i,\alpha}^{\Gamma_2}S_{j,\alpha }^{\Gamma_4} )\\
      &{}=H_{Ex}^{aa,\Gamma_1}+H_{Ex}^{aa,\Gamma_2}+H_{Ex}^{aa,\Gamma_4}+H_{Ex}^{aa,\Gamma_{12}}+H_{Ex}^{aa,\Gamma_{14}}+H_{Ex}^{aa,\Gamma_{24}}
\end{split}
\end{equation}
This equation shows the exchange interaction between spins in each state ($H_{Ex}^{aa,\Gamma_{1}}$,$H_{Ex}^{aa,\Gamma_{2}}$,$H_{Ex}^{aa,\Gamma_{4}}$) and the exchange interaction beween two states $H_{Ex}^{aa,\Gamma_{14}}$, $H_{Ex}^{aa,\Gamma_{12}}$, $H_{Ex}^{aa,\Gamma_{24}}$.

Exchange interaction between two states can be written as Eq.~\ref{eq:12} which can be proven to be zero since the final result is the dot product of the spins between different modulations:
 \begin{equation} \label{eq:12}
 \begin{split}
 H_{Ex}^{aa,\Gamma_{24}}
     &{}=\sum_{i,j}\sum_{\alpha}J_{ij}(C^2_{\alpha,a}C^4_{\alpha,a}S_{i,\alpha}^{\Gamma_2}S_{j,\alpha}^{\Gamma_4})\\ 
     &{}=\sum_{i,j}J_{ij}(F_{a,x}\bar{G}_{a,x}s_{i,x}^{\Gamma_2}s_{j,x}^{\Gamma_4}+ C_{a,y}A_{a,y}s_{i,y}^{\Gamma_2}s_{j,y}^{\Gamma_4}+\bar{G}_{a,z}F_{a,z}s_{i,z}^{\Gamma_2}s_{j,z}^{\Gamma_4})\\
     &{}=\sum_{i,j}J_{ij}(F_{a,x}\bar{G}_{a,x}(-1)^{(n_{x}^{i}+n_{y}^{i}+n_{z}^{i})}+ C_{a,y}A_{a,y}(-1)^{(n_{z}^{i})}(-1)^{(n_{x}^{i}+n_{z}^{i})}\\
     &{}+\bar{G}_{a,z}(-1)^{(n_{x}^{i}+n_{y}^{i}+n_{z}^{i})}F_{a,z})\\
     &{}=0
 \end{split}
 \end{equation}

such that we will have :
\begin{equation}
H_{Ex}^{aa,\Gamma_{12}}+H_{Ex}^{aa,\Gamma_{14}}+H_{Ex}^{aa,\Gamma_{24}}=0,
\end{equation}
and the Exchange interactions in this system can be written as follows which is the sum of the exchange interactions in each state:
\begin{equation} \label{eq:13}
\begin{split}
H^{aa}_{Ex}=H_{Ex}^{aa,\Gamma_1}+H_{Ex}^{aa,\Gamma_2}+H_{Ex}^{aa,\Gamma_4}\\
\end{split}
\end{equation}

To calculate the DMI in this system we can write:
\begin{equation} \label{eq:14}
\begin{split}
H^{aa}_{dmi}
    &{}=\sum_{i,j}\left(D_{a,ij}\times S_{j,a}\right).S_{i,a}\\
    &{}=\sum_{i,j}\left\{D_{a,ij}\times \left(C^1_{\alpha,a}S_{i,a}^{\Gamma_1}+C^2_{\alpha,a}S_{i,a}^{\Gamma_2}+C^4_{\alpha,a}S_{i,a}^{\Gamma_4}\right)\right\}.\left(C^1_{\alpha,a}S_{i,a}^{\Gamma_1}+C^2_{\alpha,a}S_{i,a}^{\Gamma_2}+C^4_{\alpha,a}S_{i,a}^{\Gamma_4}\right)\\
    &{}=H_{dmi}^{aa,\Gamma_1}+H_{dmi}^{aa,\Gamma_2}+H_{dmi}^{aa,\Gamma_4}+H_{dmi}^{aa,\Gamma_{12}}+H_{dmi}^{aa,\Gamma_{14}}+H_{dmi}^{aa,\Gamma_{24}}
\end{split}
\end{equation}
It is shown that the DMI vector in these systems can be written as Eq~.\ref{eq:15}. We are going to use this vector to describe the properties of these materials regarding DMI ~\cite{Yamaguchi1974}.
\begin{equation} \label{eq:15}
\begin{split}
D_{aa,ij}=d_{x}^a(-1)^{n_{ij}^z}\hat{i}+d_{y}^a(-1)^{n_{ij}^x+n_{ij}^y+n_{ij}^z}\hat{j}+d_{z}^a(-1)^{n_{ij}^x+n_{ij}^y}\hat{k}
\end{split}
\end{equation}

The DMI between two state can be proven to be zero since for this interaction also we will have product of the spins with two different modulations.(see Eq.~
\ref{eq:16})
\begin{equation} \label{eq:16}
\begin{split}
H_{dmi}^{aa,\Gamma_{24}}
    &{}=\sum_{i,j}\left(D_{a,ij}\times S_{j,a}^{\Gamma_2}\right).S_{i,a}^{\Gamma_4}+\left(D_{a,ij}\times S_{j,a}^{\Gamma_4}\right).S_{i,a}^{\Gamma_2}=0.
\end{split}
\end{equation}
So, the DMI between two states would be as follows:
\begin{equation} \label{eq:17}
\begin{split}
H^{aa}_{dmi}=H_{dmi}^{aa,\Gamma_1}+H_{dmi}^{aa,\Gamma_2}+H_{dmi}^{aa,\Gamma_4}.
\end{split}
\end{equation}
Hence, the total Hamiltonian of the system can be written as follows:
\begin{equation} \label{eq:18}
\begin{split}
H=H^{\Gamma_1}+H^{\Gamma_2}+H^{\Gamma_4}.
\end{split}
\end{equation}

\begin{figure}[htb!]
	\centering
	\includegraphics[width=0.5\linewidth ,keepaspectratio=true]{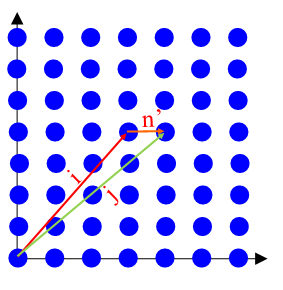} 
	\caption{Figure showing how we have considered nearest neigbours of position i.}
	\label{fig:neig}
\end{figure}

The exchange interaction in $\Gamma_{4}$ state can be written as in Eq. \ref{eq:19}. The $n'$ is the nearest neighbours at the position $i$ as shown in Fig \ref{fig:neig} and it can be ${(\pm 1,0,0)}$,${ (0,\pm 1,0)}$,${(0,0,\pm1)}$.

\begin{equation} \label{eq:19}
\begin{split}
H_{Ex}^{MM,\Gamma_4}
    &{}=\frac{1}{2}J_{M}\sum_{i}^N\sum_{j}^{NN}\bar{G}_{M,x}\bar{G}_{M,x}(-1)^{(n_{x}^{i}+n_{y}^{i}+n_{z}^{i})}(-1)^{(n_{x}^{j}+n_{y}^{j}+n_{z}^{j})}+A_{M,y}A_{M,y}(-1)^{(n_{z}^{i})}(-1)^{(n_{z}^{j})}+F_{M,z}F_{M,z}(1)^{n_{x}^{i}}(1)^{n_{x}^{j}}\\
    &{}=\frac{1}{2}J_{M}\sum_{i}^N\sum_{n'}^{NN}(\bar{G}_{M,x}^2(-1)^{(n_{x}^{i}+n_{y}^{i}+n_{z}^{i})}(-1)^{(n_{x}^{i}+n_{y}^{i}+n_{z}^{i})}(-1)^{(n'_{x}+n'_{y}+n'_{z})}\\
    &{}+(A_{M,y})^2(-1)^{(n_{z}^{i})}(-1)^{(n_{z}^{i})}(-1)^{(n'_{z})}+(F_{M,z})^2(1)^{n_{x}^{i}}(1)^{n_{x}^{i}}(1)^{n'_{x}})\\
    &{}=-3J_M(\bar{G}_{M,x})^2\sum_{i}^N(-1)^{2(n_{x}^{i}+n_{y}^{i}+n_{z}^{i})}+J_M(A_{M,y})^2\sum_{i}^N(-1)^{2(n_{z}^{i})}+3J_M(F_{M,z})^2\sum_{i}^N((1)^{2(n_{x}^{i})})\\
    &{}=-3NJ_M(\bar{G}_{M,x})^2+NJ_M(A_{M,y})^2+3NJ_M(F_{M,z})^2
\end{split}
\end{equation}

Similarly for $H_{Ex}^{a,\Gamma_1}$ and $H_{Ex}^{a,\Gamma_2}$ we can derive the following:
\begin{equation} \label{eq:20a}
\begin{split}
H_{Ex}^{MM,\Gamma_1}=NJ_M(A_{M,x})^2-3NJ_M(\bar{G}_{M,y})^2-NJ_M(C_{M,z})^2,
\end{split}
\end{equation}
\begin{equation} \label{eq:20b}
\begin{split}
H_{Ex}^{MM,\Gamma_2}=3NJ_M(F_{M,x})^2-NJ_M(C_{M,y})^2-3NJ_M(\bar{G}_{M,z})^2,
\end{split}
\end{equation} 
as well as for the exchange interactions between $R$ and $M$ sites we can write:
\begin{equation} \label{eq:21a}
\begin{split}
H_{Ex}^{RM,\Gamma_2}\simeq8NJ_{RM}F_{M,x}F_{R,x},
\end{split}
\end{equation}
\begin{equation} \label{eq:21b}
\begin{split}
H_{Ex}^{RM,\Gamma_4}\simeq8NJ_{RM}F_{M,z}F_{R,z}.
\end{split}
\end{equation}

The DMI between $M$ atoms in $H_{dmi}^{M,\Gamma_{4}}$ can be written as follows:
\begin{equation} \label{eq:22}
\begin{split}
H_{dmi}^{MM,\Gamma_4}
& =\frac{1}{2}\sum_{i}^N\sum_{n'}^{NN}\\
& \left(d_x^{M}(-1)^{(n_i^z)}(-1)^{(n'^z)}\hat{i}\times 
A_{M,y}(-1)^{(n_i^z)}(-1)^{(n'^z)}\hat{j}\right).S_{i,M}^{(\Gamma_4)}+ \\ 
& \left(d_x^{M}(-1)^{(n_i^z)}(-1)^{(n'^z)}\hat{i}\times F_{M,z}\hat{k}\right).S_{i,M}^{(\Gamma_4)}+\\
& \left(d_y^{M}(-1)^{(n_i^x+n_i^y+n_i^z)}(-1)^{(n'^x+n'^y+n'^z)}\hat{j}\times 
\bar{G}_{M,x}(-1)^{(n_i^x+n_i^y+n_i^z)}(-1)^{(n'^x+n'^y+n'^z)}\hat{i}\right).S_{i,M}^{(\Gamma_4)}+ \\ 
& \left(d_y^{M}(-1)^{(n_i^x+n_i^y+n_i^z)}(-1)^{(n'^x+n'^y+n'^z)}\hat{j}\times F_{M,z}\hat{k}\right).S_{i,M}^{(\Gamma_4)}+\\
& \left(d_z^{M}(-1)^{(n_i^x+n_i^y)}(-1)^{(n'^x+n'^y)}\hat{k}\times
\bar{G}_{M,x}(-1)^{(n_i^x+n_i^y+n_i^z)}(-1)^{(n'^x+n'^y+n'^z)}\hat{i}\right).S_{i,M}^{(\Gamma_4)}+ \\
& \left(d_z^{M}(-1)^{(n_i^x+n_i^y)}(-1)^{(n'^x+n'^y)}\hat{k}\times
A_{M,y}(-1)^{(n_i^z)}(-1)^{(n'^z)}\hat{j}\right).S_{i,M}^{(\Gamma_4)},
\end{split}
\end{equation}
and from this we can write:
\begin{equation} \label{eq:23}
\begin{split}
H_{dmi}^{MM,\Gamma_4}&=\frac{1}{2}\sum_{i}^N\\
&-12d_x^{M}(-1)^{2(n_i^z)}A_{M,y}F_{M,z}-12d_y^{M}(-1)^{2(n_i^x+n_i^y+n_i^z)}\bar{G}_{M,z}\\
&-12d_z^{M}(-1)^{2n_i^z}\bar{G}_{M,x}A_{M,y}\\
&=-6Nd_x^{M}A_{M,y}F_{M,z}-6Nd_y^{M}\bar{G}_{M,x}F_{M,z}-6Nd_z^{M}\bar{G}_{M,y}A_{M,y}.
\end{split}
\end{equation}

Similarly for $H_{dmi}^{MM,\Gamma_1}$ and $H_{dmi}^{MM,\Gamma_2}$ we can write:
\begin{equation} \label{eq:24a}
\begin{split}
H_{dmi}^{MM,\Gamma_1}=-6N d_x^{M}\bar{G}_{M,y}C_{M,z}-6N d_y^{M}C_{M,z}A_{M,x}-6N d_z^{M}A_{M,x}\bar{G}_{M,y},
\end{split}
\end{equation}
 \begin{equation} \label{eq:24b}
\begin{split}
 H_{dmi}^{MM,\Gamma_2}=-6N d_x^{M}C_{M,y}\bar{G}_{M,z}-6N d_y^{M}\bar{G}_{M,x}-6N d_z^{M}F_{M,x}C_{M,y},
\end{split}
\end{equation}
and for DMI between $R$ and $M$ we can write:
\begin{equation} \label{eq:25}
\begin{split}
H_{dmi}^{RM,\Gamma_4}=\frac{1}{2}\sum_{i}^N\sum_{j}^{NN}( (D_{ij}\times S_{j,M}^{\Gamma_4}).S_{i,R}^{\Gamma_4}\\
+\frac{1}{2}\sum_{i}^N\sum_{j}^{NN} (D_{ij}\times S_{j,R}^{\Gamma_4}).S_{i,M}^{\Gamma_4}.
\end{split}
\end{equation}
The first part of this equation is the DMI from 8 M neighbours of $R$ and $R$ which can be written as (Eq.~\ref{eq:18}):
\begin{equation} \label{eq:26}
\begin{split}
&\frac{1}{2}\sum_{i}^N\sum_{j}^{NN}(D_{ij}\times S_{j,M}^{\Gamma_4}).S_{i,R}^{\Gamma_4}=\\
&\frac{1}{2}\sum_{i}^N\sum_{n'}^{8}\left(d_y^{RM}(-1)^{(n_i^x+n_i^y+n_i^z)}(-1)^{(n'^x+n'^y+n'^z)}\hat{j}\times \bar{G}_{M,x}(-1)^{(n_i^x+n_i^y+n_i^z)}(-1)^{(n'^x+n'^y+n'^z)}\hat{i}\right).S_{i,R}^{\Gamma_4}+\\
&\frac{1}{2}\sum_{i}^N\sum_{n'}^{8}\left(d_x^{RM}(-1)^{(n_i^z)}(-1)^{(n'^z)}\hat{i}\times A_{M,y}(-1)^{(n_i^z)}(-1)^{(n'^z)}\hat{j}\right).S_{i,R}^{\Gamma_4}\\
&=\frac{1}{2}\sum_{i}^N-8d_y^{RM}(-1)^{2(n_i^x+n_i^y+n_i^z)}\bar{G}_{M,x}F_{z,R}-8d_x^{RM}(-1)^{2(n_i^z)}A_{M,y}F_{z,R}\\
&= -4Nd_y^{RM}\bar{G}_{M,x}F_{z,R}-4Nd_x^{RM}A_{M,y}F_{z,R}
\end{split}
\end{equation}

The interaction of Fe with its 8 R neighbours is as follows:
\begin{equation} \label{eq:27}
\begin{split}
&\frac{1}{2}\sum_{i}^N\sum_{j}^{NN} (D_{ij}\times S_{j,R}^{\Gamma_4}).S_{i,Fe}^{\Gamma_4}=\\
&\frac{1}{2}\sum_{i}^N\sum_{n'}^{8}( d_{y}^{RM}(-1)^{n_{i}^x+n_{i}^y+n_{i}^z}\hat{j}\times F_{z,R}\hat{k}).S_{i,Fe}^{\Gamma_4}+\\
&\frac{1}{2}\sum_{i}^N\sum_{n'}^{8}( d_{x}^{RM}(-1)^{n_{i}^z}(-1)^{(n'^z)}\hat{i}\times A_{M,y}(-1)^{(n_i^z)}(-1)^{(n'^z)}\hat{j}).S_{i,Fe}^{\Gamma_4}\\
&=-\frac{1}{2}\sum_{i}^N(8 d_{y}^{RM}(-1)^{2(n_{i}^x+n_{i}^y+n_{i}^z)}F_{z,R}\bar{G}_{M,x}+8Nd_x^{RM}(-1)^{(2n_i^z)}A_{M,y}F_{z,R})\\
&  =-4 Nd_{y}^{RM}F_{z,R}\bar{G}_{M,x}-4Nd_x^{RM}A_{M,y}F_{z,R}
\end{split}
\end{equation}

Finally for the DMI between $R$ and $M$ we can write:
\begin{equation} \label{eq:28a}
\begin{split}
 H_{dmi}^{RM,\Gamma_4}=-8Nd_{x}^{RM}F_{z,R}A_{M,y}-8Nd_{y}^{RM}F_{R,z}\bar{G}_{M,x},
\end{split}
\end{equation}
and similarly for the other states we can write:
\begin{equation} \label{eq:28b}
\begin{split}
 H_{dmi}^{RM,\Gamma_1}=-8Nd_{x}^{RM}C_{R,z}\bar{G}_{M,y}-8Nd_{y}^{RM}C_{R,z}A_{M,x},
\end{split}
\end{equation}
\begin{equation} \label{eq:28c}
\begin{split}
 H_{dmi}^{RM,\Gamma_2}=-8Nd_{x}^{RM}\bar{G}_{M,z}C_{R,y}-8Nd_{y}^{RM}F_{R,x}\bar{G}_{M,z}-8N d_{z}^{RM}F_{R,x}C_{M,y}-8N d_{z}^{RM}C_{R,y}F_{M,x}.
\end{split}
\end{equation}
By summing all the energy components and by including SIA energy we can reach the final energy of each states as follows:

\begin{equation} \label{eq:29a}
\begin{split}
H^{\Gamma_1} =  &{}H_{ex}^M+H_{DMI}^M+H_{ex}^{RM}+H_{DMI}^{RM}\\
=&{} NJ^M(A_{M,x})^2-3NJ^M(\bar{G}_{M,y})^2-NJ^M(C_{M,z})^2\\
&{} -6N d_x^{M}\bar{G}_{M,y}C_{M,z}-6N d_y^{M}C_{M,z}A_{M,x}-6N d_z^{M}A_{M,x}\bar{G}_{M,y}\\
&{} -8Nd_{x}^{RM}C_{R,z}\bar{G}_{M,y}-8Nd_{y}^{RM}C_{R,z}A_{M,x}
\end{split}
\end{equation}

\begin{equation} \label{eq:29b}
\begin{split}
H^{\Gamma_2}=   &{}H_{ex}^M+H_{DMI}^M+H_{ex}^{RM}+H_{DMI}^{RM}\\
=&{}3NJ^M(F_{M,x})^2-NJ^M(C_{M,y})^2-3NJ^M(\bar{G}_{M,z})^2\\
&{} -6N d_x^{M}C_{M,y}\bar{G}_{M,z}-6N d_y^{M}\bar{G}_{M,z}F_{M,x}-6N d_z^{M}F_{M,x}C_{M,y}\\
&{} -8NJ^{RM}F_{M,x}F_{R,x}-8Nd_{x}^{RM}\bar{G}_{M,z}C_{R,y}-8Nd_{y}^{RM}F_{R,x}\bar{G}_{M,z}\\
&{} -8N d_{z}^{RM}F_{R,x}C_{M,y}-8N d_{z}^{RM}C_{R,y}F_{M,x}
\end{split}
\end{equation}

\begin{equation} \label{eq:29c}
\begin{split}
H^{\Gamma_4}= &{}H_{ex}^M+H_{DMI}^M+H_{ex}^{RM}+H_{DMI}^{RM}+H_{SIA}^R+H_{SIA}^M\\
=&{} -3NJ^M(\bar{G}_{M,x})^2+NJ^M(A_{M,y})^2+3NJ^M(F_{M,z})^2\\
&{} -6Nd_x^{M}A_{M,y}F_{M,z}-6Nd_y^{M}\bar{G}_{M,x}F_{M,z}-6Nd_z^{M}\bar{G}_{M,x}A_{M,y}\\
&{} -8NJ^{RM}F_{M,z}F_{R,z}-8Nd_{x}^{RM}F_{z,R}A_{M,y}-8Nd_{y}^{RM}F_{R,z}\bar{G}_{M,x}\\
&{} -N K^M(\bar{G}_{M,x})^2-N K^R(\bar{G}_{R,x})^2 
\end{split}
\end{equation}

\bibliography{Mag.bib}